# An effective Landau-type model of Hf$_x$Zr$_{1-x}$O$_2$ thin film - graphene nanostructure

Anna N. Morozovska[1], Maksym V. Strikha [2, 3*], Kyle P. Kelley[4], Sergei V. Kalinin[5†], and Eugene A. Eliseev[6‡]

[1] Institute of Physics, National Academy of Sciences of Ukraine,
Pr. Nauky 46, 03028 Kyiv, Ukraine,

[2] Taras Shevchenko National University of Kyiv, Faculty of Radiophysics, Electronics and Computer Systems, Pr. Akademika Hlushkova 4g, 03022 Kyiv, Ukraine,

[3] V. Lashkariov Institute of Semiconductor Physics, National Academy of Sciences of Ukraine,
Pr. Nauky 41, 03028 Kyiv, Ukraine

[4] Center for Nanophase Materials Sciences, Oak Ridge National Laboratory,
Oak Ridge, TN 37831, USA

[5] Department of Materials Science and Engineering, University of Tennessee,
Knoxville, TN, 37996, USA

[6] Institute for Problems of Materials Science, National Academy of Sciences of Ukraine,
Krjijanovskogo 3, 03142 Kyiv, Ukraine

**Abstract**

To describe the charge-polarization coupling in the nanostructure formed by a thin Hf$_x$Zr$_{1-x}$O$_2$ film with a single-layer graphene as a top electrode, we develop the "effective" Landau-Ginzburg-Devonshire model. This approach is based on the parametrization of the Landau expansion coefficients for the polar (FE) and antipolar (AFE) orderings in thin Hf$_{1-x}$Zr$_x$O$_2$ films from a limited number of polarization-field curves and hysteresis loops. The Landau expansion coefficients are nonlinearly dependent on the film thickness $h$ and Zr/[Hf+Zr] ratio x, in contrast to h-independent and linearly x-dependent expansion coefficients of a classical Landau energy. We explain the dependence of the Landau expansion coefficients by the strong nonmonotonic dependence of the polar properties on the Hf$_{1-x}$Zr$_x$O$_2$ film thickness, grain size and surface energy. The proposed Landau free energy with five "effective" expansion coefficients, which are interpolation functions of x and $h$, describes the continuous transformation of polarization dependences on applied electric field and hysteresis loop shapes induced by the changes of x and $h$ in the range 0 < x < 1 and 5 nm < $h$ < 35 nm. Using the effective free energy, we demonstrated that the polarization of Hf$_{1-x}$Zr$_x$O$_2$ films influences strongly on the graphene conductivity, and the full correlation between the distribution of polarization and charge carriers in graphene is revealed. In accordance with our modeling, the polarization of the (5 – 25) nm thick $Hf_{1-x}Zr_xO_2$ films, which are in the ferroelectric-like or antiferroelectric-like states for the chemical compositions $0.35 \leq x \leq 0.95$, determine the concentration of carriers in graphene and can control its field dependence. The result can be promising for creation of next generation Si-compatible nonvolatile memories and graphene-ferroelectric FETs, because the working voltages applied to the $Hf_{1-x}Zr_xO_2$ film (which acts as a gate) can be relatively low (less than 2 V). These low voltages are sufficient to induce the pronounced hysteresis of ferroelectric polarization in the $Hf_{1-x}Zr_xO_2$ gate, which, due to the strong electric coupling, induces the hysteresis of graphene charge.

---

\* Corresponding author: maksym.strikha@gmail.com

† Corresponding author: sergei2@utk.edu

‡ Corresponding author: eugene.a.eliseev@gmail.com

## I. INTRODUCTION

Ferroelectric memory elements are considered as a competitive technology for information storage because of their fast-switching rate and low-power consumption. However, non-volatile ferroelectric random-access memories (**FeRAMs**) and field effect transistors (**FETs**) based on conventional perovskite ferroelectrics have serious limitations of their integration with metal-oxide semiconductors (**CMOS**) used in modern Si-technology [1, 2]. The discovery of ferroelectricity and antiferroelectricity in thin films of lead-free binary hafnium ($HfO_2$) and zirconium ($ZrO_2$) oxides made them promising candidates for the FeRAMs and FETs. Due to their fully CMOS-compatibility, the binary oxides are considered as the highly perspective materials for the FeRAMs and FeFETs by the International Roadmap for Devices and Systems ([IRDS™ 2021: Beyond CMOS](#)), as well as next-generation Si-integrable materials [1].

However, the binary oxide nanomaterials reveal a complex ferroelectric-antiferroelectric behavior due to the proximity of structural phase transitions and competition between the size-induced polar phase and the non-polar bulk phase. The bulk $HfO_2$ and $ZrO_2$ are high-k dielectrics belonging to nonpolar monoclinic symmetry groups $P2_1/c$ and $P2_1/m$, respectively, and maintain the symmetry in a wide temperature range due to very high temperatures of the structural phase transitions (above 1200 K). The bulk materials do not reveal any features inherent to the paraelectric or ferroelectric soft phonon modes even under high pressures. This lack of ferroelectricity is confirmed by Raman spectra [3, 4], which are temperature-independent (below 1200 K) and pressure-independent (below (12 – 16) GPa). The ferroelectric properties of $Hf_{1-x}Zr_xO_2$ nanomaterials are highly unusual in comparison with their bulk properties and strongly different from the properties of all known ferroelectrics. The ferroelectric properties emerging in $HfO_2$ films are determined by the polar orthorhombic $Pca2_1$ phase which, however, has a higher energy compared to the bulk monoclinic $P2_1/c$ phase, that results in the ferroelectric phase metastability. The fraction of the polar phase and spontaneous polarization value in thin $Hf_{1-x}Zr_xO_2$ films strongly depend on a substrate material, annealing and deposition conditions, thickness, and dopants (Si, Lu, Al, Gd, La, Y, etc.) concentration [5, 6, 7]. Depending on all these factors, thin $Hf_{1-x}Zr_xO_2$ films can be dielectric, ferroelectric, or antiferroelectric when x varies from 0 to 1 [8, 9]. The important role of the surface and grain boundary energy, and oxygen vacancies formed under various annealing conditions, and/or light ion bombardment has also been proven theoretically [10, 11, 12] and experimentally [8, 12, 13, 14]. However, the role of the Zr/[Hf+Zr] ratio x, size and surface effects are still very poorly described.

The promising and quite feasible are the elements based on a combination of graphene (or a 2D semiconductor) [15] and "smart" substrates with additional (electromechanical, polar and/or magnetic) degrees of functionality, such as thin ferroelectric films, because ferroelectrics can play the role of a gate due to the possibilities of the graphene conductivity control by changing the direction of their spontaneous polarization [16, 17]. Multiple experimental results [18, 19, 20, 21, 22] revealed that a graphene on a ferroelectric substrate, whose spontaneous polarization can be controlled by an external electric field [23, 24], can be considered as multifunctional systems, where the ferroelectric films (mostly perovskite oxides) act as the gate on the graphene conduction [25]. The ferroelectric spontaneous polarization can induce a very high concentration of carriers in the graphene channel of the FET, namely in two orders of magnitude higher than it can be obtained using ordinary dielectric substrates (see e.g., [16] and refs. therein). Moreover, the dependence of the graphene FET conductance on the gate voltage has a hysteretic form, and this effect can be used for construction of high-frequency elements of advanced non-volatile memories [17]. Potential barriers spontaneously arise in the plane of 180-degree domain walls, forming a set of ferroelectric gates. The potential barriers are responsible for the formation of the p–n junctions in a graphene channel due to the complex interaction between the ferroelectric polarization and the structural defects at the graphene-ferroelectric interface [26, 27]. The variety of effects enables to regard the graphene on a ferroelectric substrate as a promising smart system.

Here we consider a hybrid nanostructure "single-layer graphene – thin $Hf_{1-x}Zr_xO_2$ film", as a prototype for the FET with a graphene channel and a thin $Hf_{1-x}Zr_xO_2$ film as a gate insulator. We chose the $Hf_{1-x}Zr_xO_2$ because of its full Si-compatibility and easy reversable ferroelectric polarization, which can vary, e.g., from 5 $\mu C/cm^2$ for x=0.2 to 30 $\mu C/cm^2$ for x=0.5 [8], and does not degrade but increases or "wake-up" after multiple electric field cycling [28]. Compared with many perovskite oxides, the advantage of the CMOS-compatible $Hf_{1-x}Zr_xO_2$ gate is the relatively small value of the coercive field, which requires smaller gate voltages to switch the device. The minimal voltage is limited by the fact that the graphene has its own free charge carriers (concentration $n_0 \sim 10^{15}$ m$^{-2}$ at room temperature) in addition to the gate doping, and if the ferroelectric gate doping gives concentrations close to $n_0$ or higher, the memory effect disappears. Noteworthy, we consider a model situation, assuming a perfect electric contact between the single-layer graphene and the $Hf_{1-x}Zr_xO_2$ film, as well as the absence of electrochemically active species at the $Hf_{1-x}Zr_xO_2$ interfaces, which role are studied elsewhere [29].

We use the Landau-Ginzburg-Devonshire (**LGD**) phenomenological approach and finite element modelling (**FEM**) to describe the polar and antipolar ordering in thin $Hf_{1-x}Zr_xO_2$ (**HZO**) films. Using a limited number of polarization-field curves and hysteresis loops measured for thin $Hf_{1-x}Zr_xO_2$ films capped with conducting electrodes [8], and assuming the absence of electrochemical reactions at the film interfaces, we determine effective expansion coefficients of the Landau free energy, as a way to parametrize the strongly nonlinearly dependent on the film thickness $h$ and Zr/[Hf+Zr] ratio x.

The original part of this work contains the physical description of the problem and effective LGD approach (**Section II**). We consider in detail transformations of a polarization reversal scenario and hysteresis loops induced by the changes of the chemical composition x, surface and grain boundaries influence, and thickness of HZO films (**Section III**). The "wake-up" of HZO polarization is considered in **Section IV.** The influence of the film polarization on the graphene conductivity is analyzed in **Section V. Section VI** provides conclusions of the results. **Supplemental Material** [30] contains the description of methods and numerical algorithms, and a table of HZO material parameters.

## II. LANDAU-GINZBURG-DEVONSHIRE APPROACH

The emergent behaviors in uniaxial antiferroelectric-ferroelectric (AFE-FE) thin films have been explored within the framework of the LGD approach [31]. Here we use the effective LGD approach for the characterization of the polar (FE) and antipolar (AFE) orderings dependence on the voltage applied to the thin HZO film, and for the analysis of the static and dynamic hysteresis loops. The HZO film is sandwiched between different electrodes (e.g., made of 3D highly conducting materials or/and graphene layers).

### A. The geometry of the considered problem and electrostatic equations

The geometry of the considered problem is shown in **Fig. 1(a).** The thin HZO film is in ideal electric contact with the single-layer graphene and the ideal-conducting bottom electrode. Electrochemically active species and localized charges are absent at the graphene-oxide interface. Periodic voltage $V(t)$ is applied to the bottom electrodes. The voltage can induce the polarization reversal in the film, which takes place either via the single-domain scenario (for very efficient screening by the electrodes with an effective screening length less than 0.1 nm) or via the motion of the ferroelectric domain walls (for less efficient screening by the electrodes with an effective screening length more than 0.1 nm) [16]. A potential $\phi$ of a quasi-static electric field satisfies electrostatic equations, inside the each of the layers in the considered heterostructure.

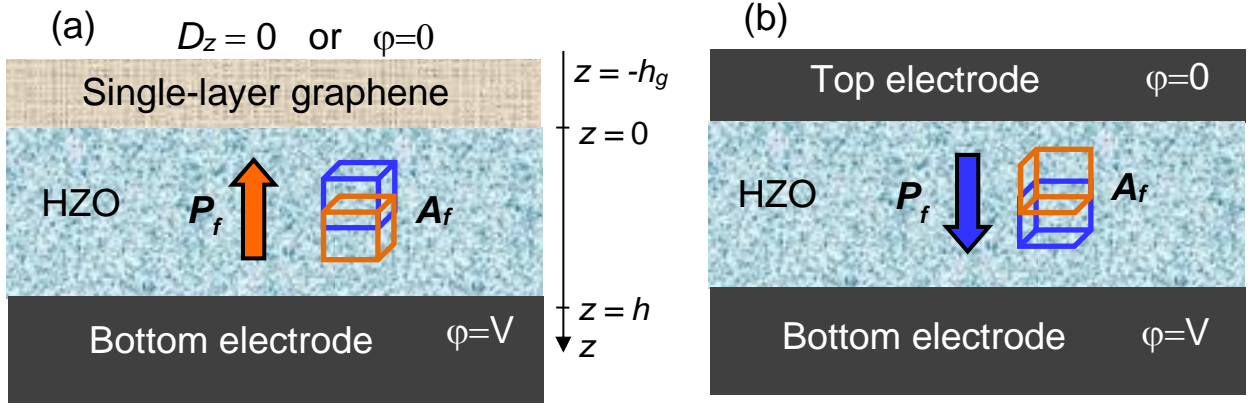

**Figure 1**. **(a)** Geometry of the considered heterostructure, consisting of the HZO film of thickness $h$ sandwiched between the single-layer graphene and conducting bottom electrode/substrate. **(b)** A HZO film sandwiched between the conducting TiN electrodes. Thick arrows show the polar order parameter, $P_f$. The antipolar order parameter, $A_f$, can be presented by microscopic dipole moments, which are counter-directed and located in the neighboring sublattices (shown by empty cubes).

**The single-layer graphene.** We consider the single-layer graphene as an ultra-thin ideally-planar sheet of thickness $h_g$, where the concentration of electrons in the conduction band and holes in the valence band are $n(\phi) = \frac{1}{h_g}\int_0^\infty d\varepsilon\, g_n(\varepsilon) f(\varepsilon - E_F - e\phi)$ and $p(\phi) = \frac{1}{h_g}\int_0^\infty d\varepsilon\, g_p(\varepsilon) f(\varepsilon + E_F + e\phi)$, respectively [16, 17]. The introduced 2D density of states, $g_n(\varepsilon) = g_p(\varepsilon) = 2\varepsilon/(\pi\hbar^2 v_F^2)$ corresponds to the gapless graphene spectrum [32], $E_F$ is the Fermi energy level and $v_F$ is the Fermi velocity in the graphene. Using the new variable, $\psi = \frac{e\phi + E_F}{k_B T}$, the graphene charge density, $\sigma_g(\psi) = e\big(p_{2D}(\psi) - n_{2D}(\psi)\big)$, is equal to [17]:

$$\sigma_g(\psi) = \frac{2(k_B T)^2 e}{\pi \hbar^2 v_F^2 h_g} \big(Li_2[-exp(\psi)] - Li_2[-exp(-\psi)]\big). \qquad (1)$$

Here $Li_n(z) = \sum_{m=1}^\infty \frac{z^m}{m^n}$ is the polylogarithm function. Corresponding Pade-exponential approximation of Eq.(1) is

$$\sigma_g(\psi) \approx \frac{2(k_B T)^2 e}{\pi \hbar^2 v_F^2 h_g}\left[\frac{1}{\eta(\psi)} - \frac{1}{\eta(-\psi)}\right], \qquad (2a)$$

where

$$\eta(\psi) = exp(\psi) + 2\left(\psi^2 + \frac{\psi}{2} + \frac{2\pi^2}{12-\pi^2}\right)^{-1}. \qquad (2b).$$

**The HZO film.** The dependence of polarization components on the inner electric field **E** can be linearized for transverse components, $P_1$ and $P_2$, as $P_1 = \varepsilon_0(\varepsilon_{11}^b - 1)E_1$ and $P_2 = \varepsilon_0(\varepsilon_{22}^b - 1)E_2$. Polarization z-component is $P_3 = P_3^f + \varepsilon_0(\varepsilon_{33}^b - 1)E_3$, where the so-called relative "background" permittivity, $\varepsilon_{ij}^b$, is introduced [33]. The values of $\varepsilon_{ij}^b$ are limited by the linear dielectric response of the lattice [34]. Evolution and spatial distribution of the ferroelectric polarization $P_3^f$ is determined from the time-dependent LGD type Euler-Lagrange equation, which will be considered in the next subsection.

Since $E_i = -\partial\phi/\partial x_i$, the coupled system of equations for the electrostatic potential $\phi$ acquires the form:

$$\varepsilon_0 \varepsilon_g \Delta \phi_g = -\sigma_g(\psi), \quad \text{for} \quad -h_g < z < 0, \tag{3a}$$

$$\varepsilon_0 \left( \varepsilon_{33}^b \frac{\partial^2}{\partial z^2} + \varepsilon_{11}^b \Delta_\perp \right) \phi_f = \frac{\partial P_3^f}{\partial z}, \quad \text{for} \quad 0 < z < h. \tag{3b}$$

Here the symbol $\Delta$ is the 3D-Laplace operator, and the symbol $\Delta_\perp$ is the 2D-Laplace operator. Boundary conditions to the system (3) are zero displacement or zero potential at the top of graphene layer ($z = -h_g$), the fixed periodic potential $V(t)$ at the bottom electrode ($z = h$), the continuity of the electric potentials, $\phi_g$ and $\phi_f$, and the continuity of the electric displacement normal components, $D_3^g$ and $D_3^f$, at the graphene layer ($z = 0$). The explicit form of the boundary conditions is:

$$\phi_g(x, y, -h_g) = 0, \quad \text{or} \quad D_3^g(x, y, -h_g) = 0, \tag{4a}$$

$$\phi_g(x, y, 0) = \phi_f(x, y, 0), \quad D_3^g(x, y, 0) = D_3^f(x, y, 0), \tag{4b}$$

$$\phi_f(x, y, h) = V(t). \tag{4c}$$

Here $D_3^f = \varepsilon_0 \varepsilon_{33}^b E_3 + P_3^f$ and $D_3^g = \varepsilon_0 \varepsilon_g E_3$, where $\varepsilon_g$ is the effective dielectric permittivity of a single-layer graphene, which value is still under debate (see e.g., Ref. [35]).

### B. The spatial-temporal evolution of polarization in the HZO film

To determine the spatial-temporal evolution of polarization in the HZO film we use the Kittel-type model [36] incorporating polar and antipolar modes [37, 38, 39] combined with the LGD approach [31]. Corresponding LGD free energy functional $F$ additively includes a bulk part – an expansion on the 2-th and 4-th powers of the polar ($P_f$) and antipolar ($A_f$) order parameters, $F_{bulk}$; a polarization gradient energy contribution, $F_{grad}$; an electrostatic contribution, $F_{el}$; and a surface energy, $F_S$. It has the form [29]:

$$F = F_{bulk} + F_{grad} + F_{el} + F_S, \tag{5}$$

where the constituent parts are

$$F_{bulk} = \int_{V_f} d^3r \left( \frac{a_P}{2} P_f^2 + \frac{b_P}{4} P_f^4 + \frac{\eta}{2} P_f^2 A_f^2 + \frac{a_A}{2} A_f^2 + \frac{b_A}{4} A_f^4 \right), \tag{6a}$$

$$F_{grad} = \int_{V_f} d^3r \frac{g_{ij}}{2} \left( \frac{\partial P_f}{\partial x_i} \frac{\partial P_f}{\partial x_j} + \frac{\partial A_f}{\partial x_i} \frac{\partial A_f}{\partial x_j} \right), \tag{6b}$$

$$F_{el} = -\int_{V_f} d^3r \left( P_i E_i + \frac{\varepsilon_0 \varepsilon_b}{2} E_i E_i \right), \tag{6c}$$

$$F_S = \frac{1}{2} \int_S d^2r \left( c_P P_f^2 + c_A A_f^2 \right). \tag{6d}$$

Here $V_f$ is the film volume and $S$ is the film surface area.

It is important to consider the dependencies of LGD expansion coefficients on external and internal factors, such as temperature, sizes, elastic stresses and/or strains. Indeed, the role of elastic stresses and strains, such as Vegard (or chemical) stresses [11] and/or misfit strains originated from the film-substrate lattice mismatch [40], can be very important. The electrostriction couples the elastic fields with electric polarization and antipolar ordering. As a rule, homogeneous strains and/or stresses, which constitute the elastic energy, lead to the renormalization of LGD expansion coefficients in thin ferroelectric [41, 42] and antiferroelectric [31] films. As a result the strain-induced phase transitions and phases absent in the bulk often occur in thin films [31, 40-42]. Since the LGD coefficients in Eqs.(6), as well as electrostriction coefficients of HZO, are poorly known, we regarded that they are already renormalized by homogeneous elastic strains. The analytical description of inhomogeneous elastic stresses, which likely exist in HZO films, are much more complex, because their dependence on the grain size is unknown.

For classical ferroelectric films with a pronounced temperature-dependent and strain-dependent soft mode, the coefficients $a_P$ and $a_A$ linearly depend on the temperature and strains induced by the film-substrate lattice mismatch (see e.g., Refs. [41,42]). In comparison, for HZO films we assume a general-form dependence of $a_P$ and $a_A$ on the chemical composition $x$, average grain radius $R$ and film thickness $h$:

$$a_P(x,R,h) = \xi_P(x) + \Delta_P(x,R,h), \tag{7a}$$
$$a_A(x,R,h) = \xi_A(x) + \Delta_A(x,R,h). \tag{7b}$$

The "bulk" coefficients $\xi_P(x)$ and $\xi_A(x)$ are positive or zero in accordance with the absence of polar and antipolar modes in the homogeneous bulk $HfO_2$ and $ZrO_2$.

The second terms in Eqs.(7) occur in thin films due to several reasons, such as the surface reconstruction, homogeneous elastic strains, grain boundaries energy and/or chemical-type Vegard stresses. Since the energy excess of a columnar grain is [8]:

$$G_{grain}(x,R,h) = G_{bulk}(x) + \frac{1}{\pi R^2 h}\left(2\pi R^2 \gamma_{if} + 2\pi R h \gamma_{gb}\right), \tag{8a}$$

one can assume that

$$\Delta_P(x,R,h) \cong Q_{3i}(x)\left(\frac{2\gamma_i^{if}(x)}{h} + \frac{2\gamma_i^{gb}(x)}{R}\right) + Q_{3i}(x)\sigma_i(R,h), \tag{8b}$$

$$\Delta_A(x,R,h) \cong Z_{3i}(x)\left(\frac{2\gamma_i^{if}(x)}{h} + \frac{2\gamma_i^{gb}(x)}{R}\right) + Z_{3i}(x)\sigma_i(R,h). \tag{8c}$$

Here $\gamma_{if}$ is the interface energy and $\gamma_{gb}$ is the grain boundary energy per unit surface, which, as a rule, are anisotropic in crystalline solids; $Q_{ij}$ and $Z_{ij}$ are the components of electrostriction tensors, and $\sigma_i$ are elastic stresses, written in Voigt notations, which also depend on $x$, $R$ and $h$. Due to the strong dependence of the orthorhombic phase energy on $x$, $R$ and $h$, thin HZO films can be either dielectric as the bulk, or reveal ferroelectric-like or antiferroelectric-like behaviors when $x$ varies from 0 to 1 and $h$ varies from e.g., 10 nm to 30 nm [8, 9]. Since the average grain size $R$ depends on the film preparation conditions, $x$ and $h$, the dependences of $a_P(x,h)$ and $a_A(x,h)$ on $x$ and $h$ can be determined from experimental results. Below we use the results of Park et al [8] for the purpose.

The conditions $b_P > 0$, $b_A > 0$, and $\eta > -\sqrt{b_P b_A}$ should be valid for the functional stability; they also correspond to the absence of polar and antipolar modes in bulk materials $HfO_2$ and $ZrO_2$. Also, the temperature-independent matrix of gradient coefficients $g_{ij}$ should be positively defined for the free energy stability.

Spatial-temporal evolution of the polar and antipolar order parameters, $P_f$ and $A_f$, is determined from the coupled time-dependent LGD type Euler-Lagrange equations, obtained by the minimization of the free energy $F$:

$$\Gamma_P \frac{\partial P_f}{\partial t} + a_P P_f + b_P P_f^3 + \eta A_f^2 P_f - g \Delta P_f = E_3, \qquad (9a)$$

$$\Gamma_A \frac{\partial A_f}{\partial t} + a_A A_f + b_A A_f^3 + \eta P_f^2 A_f - g \Delta A_f = 0. \qquad (9b)$$

$\Gamma_P$ and $\Gamma_A$ are Landau-Khalatnikov relaxation coefficients [43], and $g$ is a positive gradient coefficient written in an isotropic approximation. The relaxation times of $P_f$ and $A_f$ are $\tau_P = \Gamma_P/|a_P|$ and $\tau_A = \Gamma_A/|a_A|$, respectively. Corresponding boundary conditions to Eqs.(9) are of the third kind [44]:

$$\left(P_f - \Lambda_P \frac{\partial P_f}{\partial z}\right)\bigg|_{z=0} = 0, \quad \left(P_f + \Lambda_P \frac{\partial P_f}{\partial z}\right)\bigg|_{z=h} = 0, \qquad (10a)$$

$$\left(A_f - \Lambda_A \frac{\partial A_f}{\partial z}\right)\bigg|_{z=0} = 0, \quad \left(A_f + \Lambda_A \frac{\partial A_f}{\partial z}\right)\bigg|_{z=h} = 0. \qquad (10b)$$

Here $\Lambda_P = \frac{g}{c_P}$ and $\Lambda_A = \frac{g}{c_A}$ are extrapolation lengths, which physical range is (0.5 – 5) nm [45]. As a rule, the polar order parameter is observable (i.e., measurable), and the antipolar order parameter cannot be directly measured. However, the nonlinear coupling between $P_f$ and $A_f$ changes the field dependence $P_f(E_3)$. Below we consider the HZO film at room temperature. The temperature dependence of the polar properties of doped $HfO_2$ is considered elsewhere [46].

### III. POLARIZATION REVERSAL IN THE PERFECTLY SCREENED HZO FILM

To establish the baseline physics of the HZO film associated with extrinsic size effects, at first we consider the case of ideal screening by the electrodes. The graphene layer is absent for the case, as shown in **Fig. 1(b)**, and the perfect electric contact between the HZO film and conducting electrodes provides a full screening of the out-of-plane polarization component and prevents the domain formation. We also apply the so-called "natural" boundary conditions, $\frac{\partial A_f}{\partial z}\big|_{z=0,h} = 0$ and $\frac{\partial P_f}{\partial z}\big|_{z=0,h} = 0$, since they prevent the surface-induced polarization inhomogeneity.

The thermodynamically stable phases of the free energy (5) are the spatially homogeneous dielectric (**DE**), paraelectric (**PE**), ferroelectric-type polar (**FE**), mixed ferrielectric (**FI**) and antiferroelectric-type antipolar (**AFE**) phases (see **Fig. 2**). Corresponding values of the spontaneous order parameters, free energy densities, and stability conditions of these phases are listed in **Table I**. The order parameters $A_f$ and $P_f$, are decoupled for $\eta = 0$, and thus the phase diagram, shown in **Fig. 2(a)** consists of four equal quadrants, which intersect along the lines $a_P = 0$ and $a_A = 0$. The DE+PE phases occupies the first quadrant, $a_P > 0$ and $a_A > 0$, at that the PE phase is located near the vertical line $a_P = 0$. For the case $\eta = 0$ the FE, FI, and AFE phases occupy the second, the third and the fourth quadrants, respectively. For the case $0 < \eta \leq 1$ the region of the FI phase gradually decreases with $\eta$ increase, being substituted by the FE and AFE phases [see **Figs. 2(b)** and **2(c)**], and eventually disappears for $\eta > 1$ [see **Fig. 2(d)**]. However, for $\eta > 1$, the coexistence region, where the FE phase is stable and the AFE phase is metastable (or vice versa) appear in the third quadrant, $a_P < 0$ and $a_A < 0$ [see small color images in **Figs. 2(d)**].

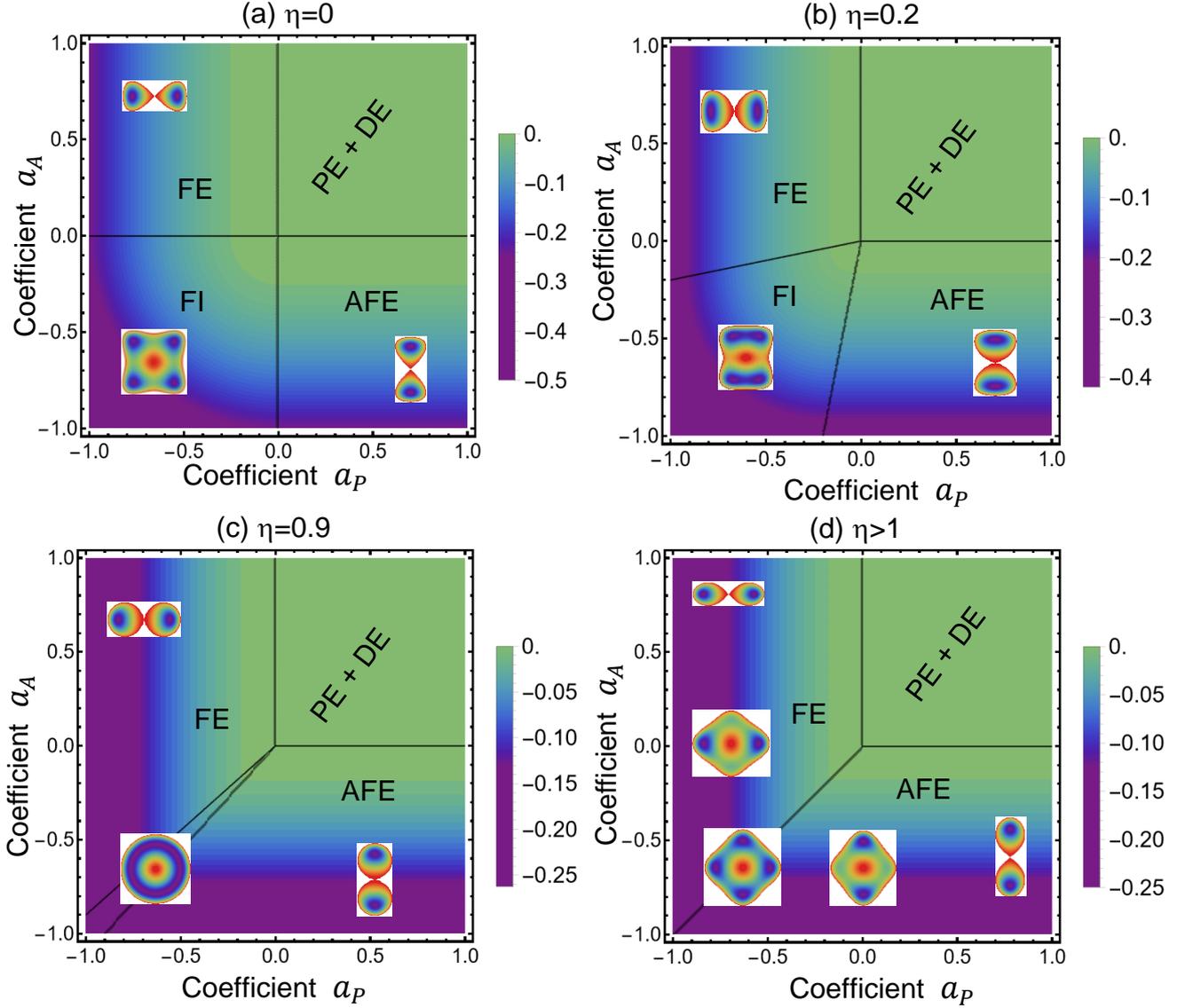

**Figure 2.** The free energy minimum in dependence on the coefficients $a_P$ and $a_A$ calculated for $b_P = b_A = 1$, $\eta = 0$ (**a**), $\eta = 0.2$ (**b**), $\eta = 0.9$ (**c**) and $\eta > 1$ (**d**). Color scales show the free energy minimum in dimensionless units. The regions of the DE + PE, FE, FI and AFE phases are separated by thin black lines. Small color images are the free energy relief in $\{P_f, A_f\}$ coordinates.

**Table I.** The spontaneous order parameters, free energy densities, stability conditions of the thermodynamically stable spatially homogeneous phases of the free energy (5), and critical field(s)

| Phase and type of hysteresis loops | Spontaneous values of the order parameters | Free energy density and stability conditions | Coercive and/or critical field(s) $E_c$, and/or $P(E_c)$ * |
|---|---|---|---|
| Dielectric (DE) and paraelectric (PE) phases. Linear (DC) and paraelectric-type (PC) curves ** | $P_f = A_f = 0$ | $f_D = 0$ <br> $a_P > 0, a_A > 0$ | absent |
| Polar phase (FE) Ferroelectric-type single loops (SL) | $A_f = 0,$ <br> $P_f = \pm\sqrt{-\frac{a_P}{b_P}},$ <br> $P_r \approx P_f$ | $f_P = -\frac{a_P^2}{4b_P}$ <br> $f_P = min, a_P < 0,$ <br> $a_A b_P - \eta a_P > 0$ | $E_c = \pm\frac{2}{3\sqrt{3}}\frac{(-a_P)^{3/2}}{b_P}$ <br> $P_f(E_c) = 0$ |
| Mixed phase (FI) Pinched loops (PL) | $A_f = \pm\sqrt{-\frac{a_A b_P - \eta a_P}{b_A b_P - \eta^2}}$ <br> $P_f = \pm\sqrt{-\frac{a_P b_A - \eta a_A}{b_A b_P - \eta^2}}$ | $f_{PA} = \frac{-b_P a_A^2 - b_A a_P^2 + \eta a_P a_A}{4(b_A b_P - \eta^2)}$ <br> $f_{PA} = min,$ <br> $a_A b_P - \eta a_P < 0, a_P b_A - \eta a_A < 0, b_A b_P > \eta^2$ | $E_c = \frac{\pm 2}{3\sqrt{3}}\frac{\left(-a_P + \frac{\eta a_A}{b_A}\right)^{3/2}}{b_P - \frac{\eta^2}{b_A}}$ <br> $P_f(E_c) = 0$ |
| Antipolar phase (AFE) Antiferroelectric-type double loops (DL) | $A_f = \pm\sqrt{-\frac{a_A}{b_A}},$ <br> $P_f = 0.$ | $f_A = -\frac{a_A^2}{4b_A}$ <br> $f_A = min, a_A < 0,$ <br> $a_P b_A - \eta a_A > 0,$ <br> $\eta > \sqrt{b_A b_P}$ | $E_{c1} = \pm\left(a_P - \frac{a_A}{\eta}b_P\right)\sqrt{-\frac{a_A}{\eta}},$ <br> $P_{c1}(E_{c1}) = \pm\sqrt{-\frac{a_A}{\eta}}$ <br> $E_{c2} = \frac{\pm 2}{3\sqrt{3}}\frac{\left(a_P - \frac{\eta a_A}{b_A}\right)^{3/2}}{\frac{\eta^2}{b_A} - b_P}$ <br> $P_{c2}(E_{c2}) = \pm\sqrt{\frac{a_P b_A - \eta a_A}{3(\eta^2 - b_P b_A)}}.$ |

\* See **Fig. 3** for designations of $P_{ci}$ and $E_{ci}$;
\*\* Noteworthy that very thin, slim and/or banana-shaped [47] dynamic polarization loops correspond to DC and PC static curves, respectively.

When the electric voltage $V$ is applied between the electrodes, the polarization reversal scenario depends on the phase. Namely, a hysteresis effect is absent for the dielectric-type (**DC**) and paraelectric-type (**PC**) curves in the DE and PE phases. Ferroelectric-type single loops (**SL**) exist in the FE phase, pinched loops (**PL**) are stable in the FI phase (as well as PL may exist in the AFE-FE coexistence region), and antiferroelectric-type double loops (**DL**) of polarization hysteresis exist in the AFE phase.

The polarization-field dependences, $P_f(E)$, shown in **Fig. 3**, correspond to the Hf$_{1-x}$Zr$_x$O$_2$ films with the ratio $x = 0.00, 0.20, 0.45, 0.70, 1.00$ and thickness $h = 10, 20$ and $30$ nm, which are sandwiched between the ideally conducting electrodes. Dimensionless parameters $b_P = b_A = 1$ and $\eta = 2$, other parameters are listed in **Table II**. These parameters were selected using the trends shown in **Fig. 2**. Polarization is normalized on the experimentally determined spontaneous value, $P_S \cong 0.15$ C/m$^2$, and electric field is normalized on the coercive field value, $E_C \cong 0.15$ V/nm. The dimensionless frequency $w$ is the frequency ω (in s$^{-1}$) multiplied on $\tau_P$, $w = \omega\tau_P$, and we regard that $\tau_P = \tau_A$ in Eqs.(9). The difference between the dynamic solid loops, calculated for the frequency $w = 0.1$, and the static dashed curves, calculated for $w = 0$, are relatively small for the DCs and PCs, and relatively big for the SLs, PLs and DLs.

The shape of the polarization loops, $P_f(E)$, shown in **Fig. 3**, are very similar to the experimental dependences, $P_{exp}(E)$, shown in Fig. 1 in Ref. [8]. The dependences $P_{exp}(E)$ were measured by Park et al [8] in Hf$_{1-x}$Zr$_x$O$_2$ films with $x = 0.00, 0.19, 0.43, 0.70$ and $1.00$ for $h = 9.2$ nm, 19.2 and 29.2 nm. The films were sandwiched between conducting TiN electrodes.

Notice, that the thinnest 10-nm HZO film reveals the DE and PE type behavior, namely the DCs, for x ≤ 0.3. The FE type SLs appear for 0.35 ≤ x ≤ 0.55, and the AFE type DLs exist for 0.7 ≤ x ≤ 0.9. The DCs appear again for x > 0.95 (see the bottom row in **Fig. 3**). The 20-nm HZO film reveals the DE and PE type behavior (namely the DCs) for x ≤ 0.35. The SLs of the FE type become stable for 0.4 ≤ x ≤ 0.55. The "slim" AFE-type DLs exist in the x-range 0.7 ≤ x ≤ 1 (see the middle row in **Fig. 3**). The thickest 30-nm HZO film reveals the DE and PE type behavior for x ≤ 0.4, and the AFE-type DLs are stable in the range 0.45 ≤ x ≤ 0.90. The DLs gradually transform to the PCs for x > 0.95 (see the top row in **Fig. 3**).

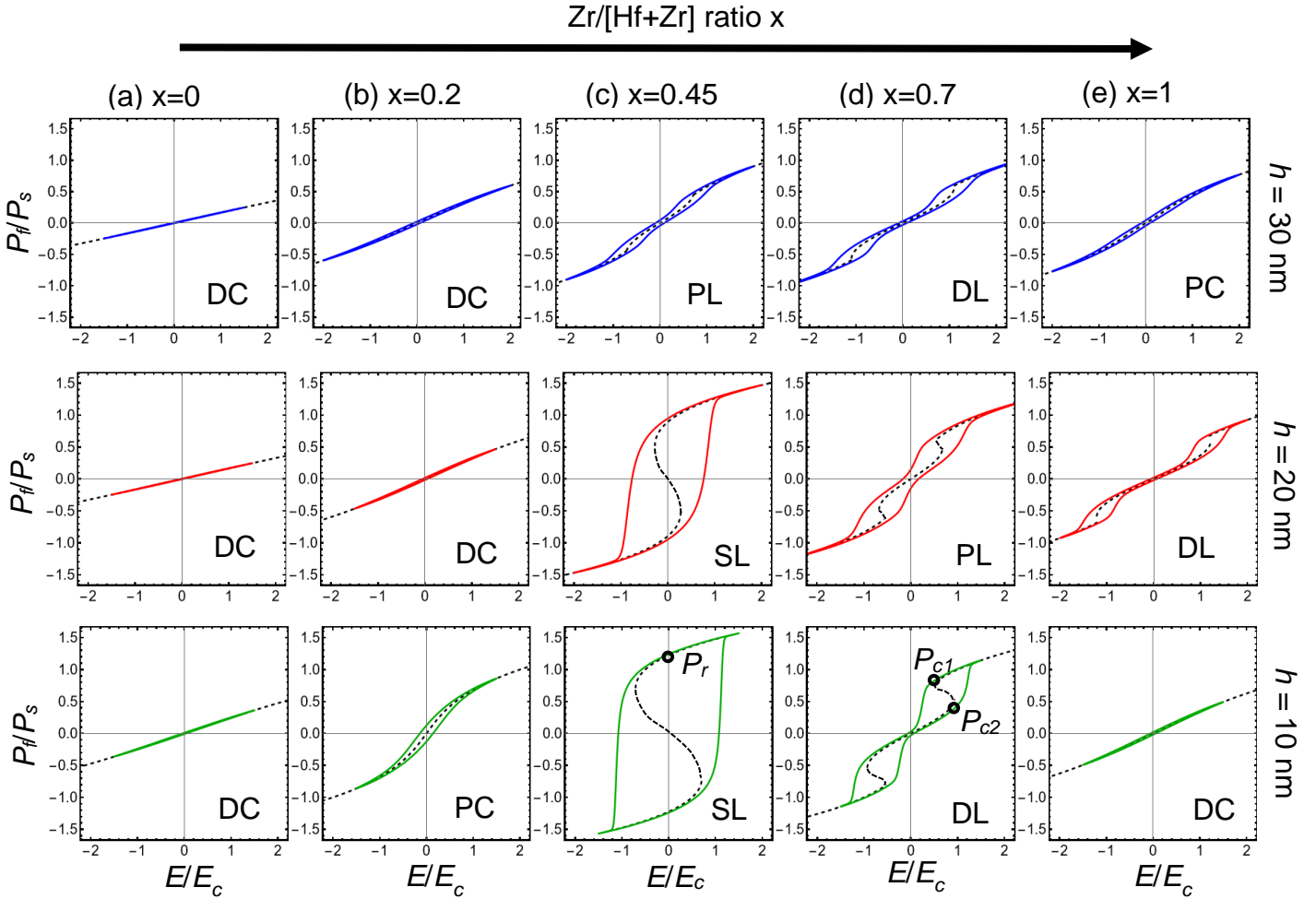

**Figure 3.** The field dependences of polarization, $P_f(E)$, calculated for Hf$_{1-x}$Zr$_x$O$_2$ films with $x = 0.00$ **(a)**, 0.20 **(b)**, 0.45 **(c)**, 0.70 **(d)**, 1.00 **(e)**, thickness $h = 10$ nm (the bottom row), 20 nm (the middle row) and 30 nm (the top row). Polarization $P_f$ is normalized on the spontaneous value, $P_S = 0.15$ C/m$^2$, and electric field $E$ is normalized on the coercive field value, $E_C = 0.15$ V/nm. Solid loops and curves correspond to the dimensionless frequency $w = 0.1$, and dashed curves are static dependences ($w = 0$). Dimensionless parameters $b_P = b_A = 1$, $\eta = 2$. Other parameters are listed in **Table II**.

**Table II.** Dimensionless parameters, $a_P$ and $a_A$, of thin Hf$_{1-x}$Zr$_x$O$_2$ films

| Thickness $h$ | $h = 10$ nm | | $h = 20$ nm | | $h = 30$ nm | |
|---|---|---|---|---|---|---|
| Content $x$ | $a_P$ | $a_A$ | $a_P$ | $a_A$ | $a_P$ | $a_A$ |
| 0 | 4.0 | 4.0 | 6.0 | 4.0 | 6.0 | 4.0 |
| 0.2 | 1.0 | 2.0 | 3.0 | 2.0 | 3.0 | 2.0 |
| 0.45 | -1.5 | 1.0 | -0.8 | 1.0 | 1.4 | -0.4 |
| 0.70 | 0.0 | -1.3 | 0.5 | -0.8 | 1.5 | -0.7 |
| 1.00 | 3.0 | 1.0 | 1.3 | -0.9 | 2.0 | -0.1 |

We chose LGD parameters, listed in **Table II**, to reach qualitative agreement with the experimental results [8]. The parameters allow a description of the long-range polar and/or antipolar order in HZO films of discrete thickness (10, 20 and 30 nm) and Zr/[Hf+Zr] ratios (0, 0.2, 0.45, 0.7 and 1). We further proceed to the qualitative determination of the LGD-model parameters in a continuous range of $x$ and $h$, namely for 5 nm $< h <$ 30 nm and $0 \le x \le 1$. First, we need to select the interpolation functions for the coefficients $a_P(x, h)$ and $a_A(x, h)$. Here, from Eqs.(7) and (8) the coefficients $a_P(x, h)$ and $a_A(x, h)$ are complicated functions of $R$, $x$ and $h$, and the conditions $a_P(x, h) < 0$ and/or $a_A(x, h) < 0$ are required for the stability of FE, FI or AFE phases (see **Tables I** and **II**). Since the average grain size $R$ is the function of $x$ and $h$, and because the polar and/or antipolar orders exist in the definite range of HZO thickness, $h_{min} < h < h_{max}$, and in the definite x-range, $x_{min} < x < x_{max}$, we can suggest that the conditions $a_P(x, h) < 0$ and/or $a_A(x, h) < 0$ are valid in the range $h_{min} < h < h_{max}$. Consequently, $a_P(x, h) > 0$ and $a_A(x, h) > 0$ outside the ranges. For instance, the $h_{min}$ is about 9 nm and $h_{max}$ is about 29 nm; $x_{min}$ is about 0.2 and $x_{max}$ is about (0.9 – 1) in accordance with the experiment [8].

Using the Taylor series expansion near the minima of unknown (but analytic) functions, the simplest parabolic approximations for the dependences of the coefficients $a_P(x, h)$ and $a_A(x, h)$ on $h$ can be assumed:

$$a_P(x, h) = a_{P0} + 4\alpha_P(x) \frac{(h - h_{min}^P)(h_{max}^P - h)}{(h_{max}^P)^2 - (h_{min}^P)^2}, \tag{11a}$$

$$a_A(x, h) = a_{A0} + 4\alpha_A(x) \frac{(h - h_{min}^A)(h_{max}^A - h)}{(h_{max}^A)^2 - (h_{min}^A)^2}. \tag{11b}$$

Here $a_{P0}$ and $a_{A0}$ are the "aristo-phase" coefficients, which are small, positive, h- and x-independent. Here the term "aristo-phase" means a hypothetic parent phase of higher symmetry without any long-range order (e.g., in classical ferroelectric perovskites aristo-phase is a real high-temperature cubic m3m phase).

The x-dependent prefactors $\alpha_P(x)$ and $\alpha_A(x)$ are assumed negative. Therefore, the size-dependent part of the coefficient $a_P(x, h)$ is negative for $h_{min}^P < h < h_{max}^P$ and the polar order exists in the thickness range. The size-dependent part of the coefficient the coefficient $a_A(x, h)$ is negative for $h_{min}^A < h < h_{max}^A$ and the antipolar order exists in the thickness range. The thicknesses, $h_{min}^P$, $h_{max}^P$, $h_{min}^A$ and $h_{max}^A$, are fitting parameters, which can be determined from the thickness ranges of the FE, FI and AFE phases existence in thin HZO films.

For illustrative purposes, below one can assume the Gaussian-type x-dependence of the prefactors $\alpha_P(x)$ and $\alpha_A(x)$ in the coefficients $a_P(x, h)$ and $a_A(x, h)$, namely:

$$\alpha_P(x) = \alpha_{P0}\text{Exp}\left[-\frac{(x-x_P)^2}{(\Delta x_P)^2}\right], \tag{11c}$$

$$\alpha_A(x) = \alpha_{A0}\text{Exp}\left[-\frac{(x-x_A)^2}{(\Delta x_A)^2}\right]. \tag{11d}$$

The negative parameters $\alpha_{P0}$ and $\alpha_{A0}$, the x values, $x_P$ and $x_A$, and the dispersions, $\Delta x_P$ and $\Delta x_A$, are fitting parameters, which can be determined from the x-range of the FE, FI and AFE phases existence in thin HZO films. This assumption is based on hypothesis that the FE and AFE properties of thin HZO films exist in the limited range of Hf or Zr composition and are the most pronounced near the compositions $x_P$ and $x_A$, respectively. If the experimentally observed FE or AFE properties, such as the spontaneous polarization and/or coercive fields values, are well-localized in the x-range and have a pronounced maximum, the Gaussian-type fitting functions (11c) – (11d) with small dispersions $\Delta x_P$ and $\Delta x_A$ are well-suitable for the description of such experimental results. Otherwise, non-Gaussian fitting functions (e.g., step-like functions) should be used. Below we use the Gaussian functions (11c) – (11d) for qualitative description of the experimental results [8].

The $x$ and $h$ dependences of the dimensionless LGD-coefficients, $a_P(x, h)$ and $a_A(x, h)$, calculated from Eqs.(11), are shown in **Figs. 4(a)** and **4(b)**, respectively. **Figure 4(c)** is the $x - h$ contour map, which contains the blue, light grey-green and light green regions of the FE, FI and AFE phases, where the SL, PL and DL loops exist, respectively. The contours of the FE, FI and AFE regions are the boundaries between different types of hysteresis loops.

The relatively deep potential wells, which belong to the FE and AFE states, are separated by the energy barrier and surrounded by a zero-energy region of the DE + PE phase [see **Fig. 4(d)**]. The DC and PC correspond to the DE and PE phase states, respectively. The contour map in **Fig. 4(c)** is in a qualitative agreement with the experimental phase diagram shown in Fig. 7 in Ref. [8], which proofs that the usage of parabolic approximations (11a) – (11b) and Gaussian functions (11c) – (11d) as the fitting functions is grounded in the considered case. However, for quantitative description of the experimental results, the fitting of experimentally measured dielectric permittivity and polarization hysteresis loops is required, and the possibility is demonstrated in **Appendix A** [30].

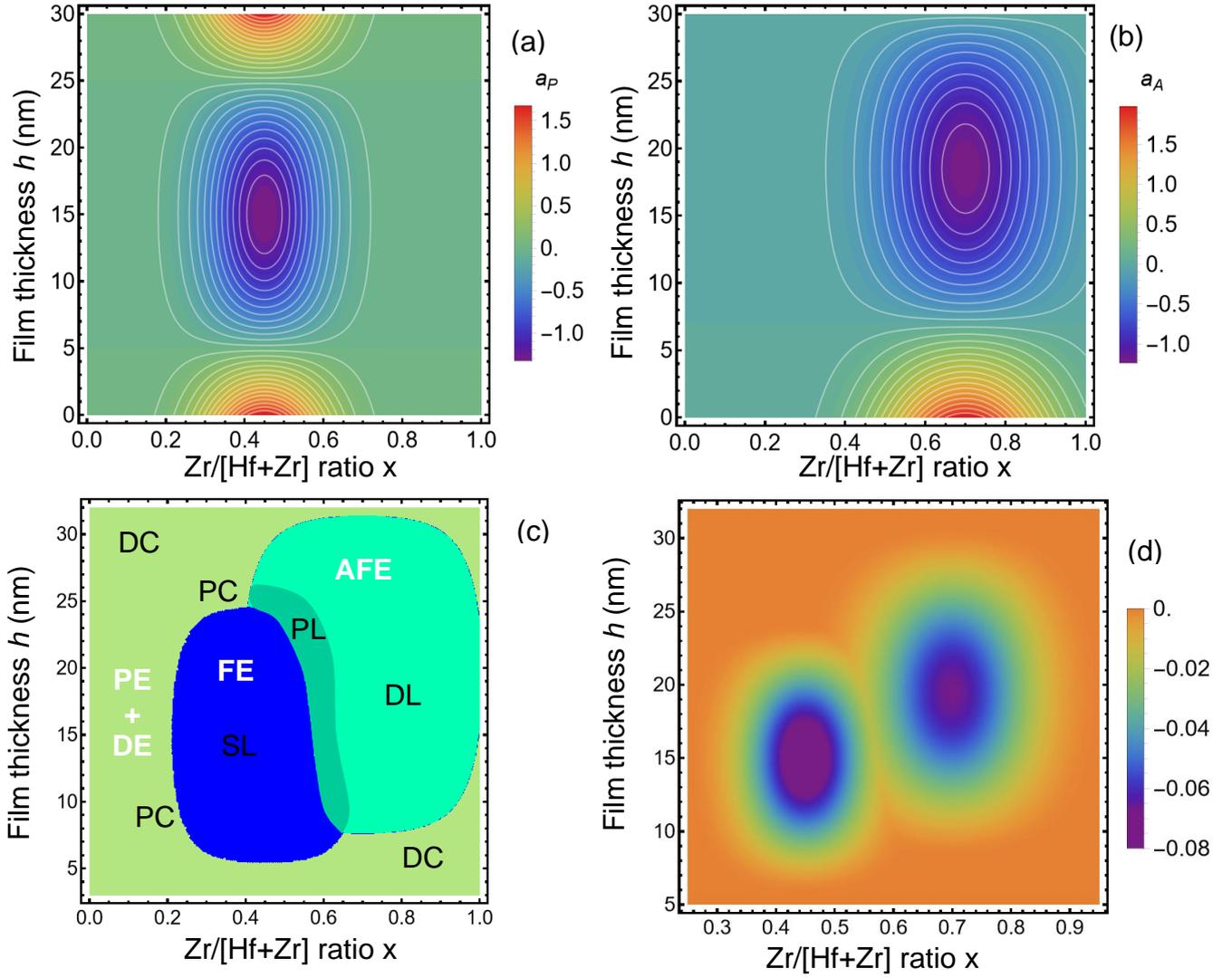

**Figure 4.** The dimensionless coefficients $a_P(x, h)$ (a) and $a_A(x, h)$ (b) in dependence on $x$ and $h$. (c) The diagram of the PE, FE, FI and AFE phase regions in the $x - h$ coordinates. The contours are for the boundaries of the FE (the blue region of SLs), FI (the light grey-green region of PLs) and AFE (the light green region of DLs) phases, respectively. (d) The free energy minimum in dependence on $x$ and $h$. The color scale shows the energy minimum in dimensionless units. Parameters $\alpha_{P0} = -2$, $\Delta x_P = 0.15$, $x_P = 0.45$, $h_{min}^P = 5$, $h_{max}^P = 25$, $\alpha_{A0} = -2$, $\Delta x_A = 0.20$, $x_A = 0.70$, $h_{min}^A = 7$, and $h_{max}^A = 32$, $b_P = b_A = 1$ and $\eta = 2$.

Noteworthy, that we also determined the "effective" Landau coefficients, $a_P(x, h)$, $a_A(x, h)$, $b_P$, $b_A$ and $\eta$, in the physical units from the quantitative comparison with the experimental results [8]. The algorithm is described in detail in **Appendix A** [30] and schematically shown in **Fig. 5**. In brief, we used a limited number (25) of polarization curves and hysteresis loops, $P_{exp}(E)$, measured for thin HZO films capped with conducting TiN electrodes, which are shown in Fig. 1 in Ref. [8]. Red, orange, yellow, cyan, blue and green circles in **Fig. 5(a)** and **5(b)** correspond to different types of polarization curves and loops, which were experimentally observed for 25 pairs $\{x_i, h_j\}$ ($i, j = 1, 2, ..., 5$) and shown in Fig. 7 in Ref. [8]. The nine dielectric and two paraelectric curves are presented by red and orange circles, respectively. A very thin paraelectric-like loop is shown by a yellow circle. Two thin ferroelectric-like loops and three wide ferroelectric-type loops are presented by light cyan and dark blue circles, respectively; and eight antiferroelectric-type double hysteresis loops are presented by green circles. The 25 pairs $\{x_i, h_j\}$ are composed from $x_i = 0, 0.19, 0.43, 0.7, 1$ and $h_j = 9.2, 14.2, 19.2, 24.2, 29.2$ nm in accordance with Fig. 7 in Ref. [8].

Since the TiN electrodes are conducing, we assume a perfect screening of HZO polarization in the electrodes, and the absence of surface electrochemical effects at the interfaces. Using these assumptions, we determine the effective expansion coefficients of the Landau free energy (6a) from the fitting of the remanent polarization for SLs or PLs, and parameters of the DL openings, linear dielectric susceptibility and loop shape using theoretical dependencies $P_f(E)$. The effective coefficients $a_P(x, h)$ and $a_A(x, h)$ are strongly nonlinearly dependent on the film thickness $h$ and Zr/[Hf+Zr] ratio x, in contrast to $h$-independent and linearly x-dependent expansion coefficients of a classical Landau energy describing FE-AFE solid solutions, such as a bulk $Pb_{1-x}Zr_xO_3$.

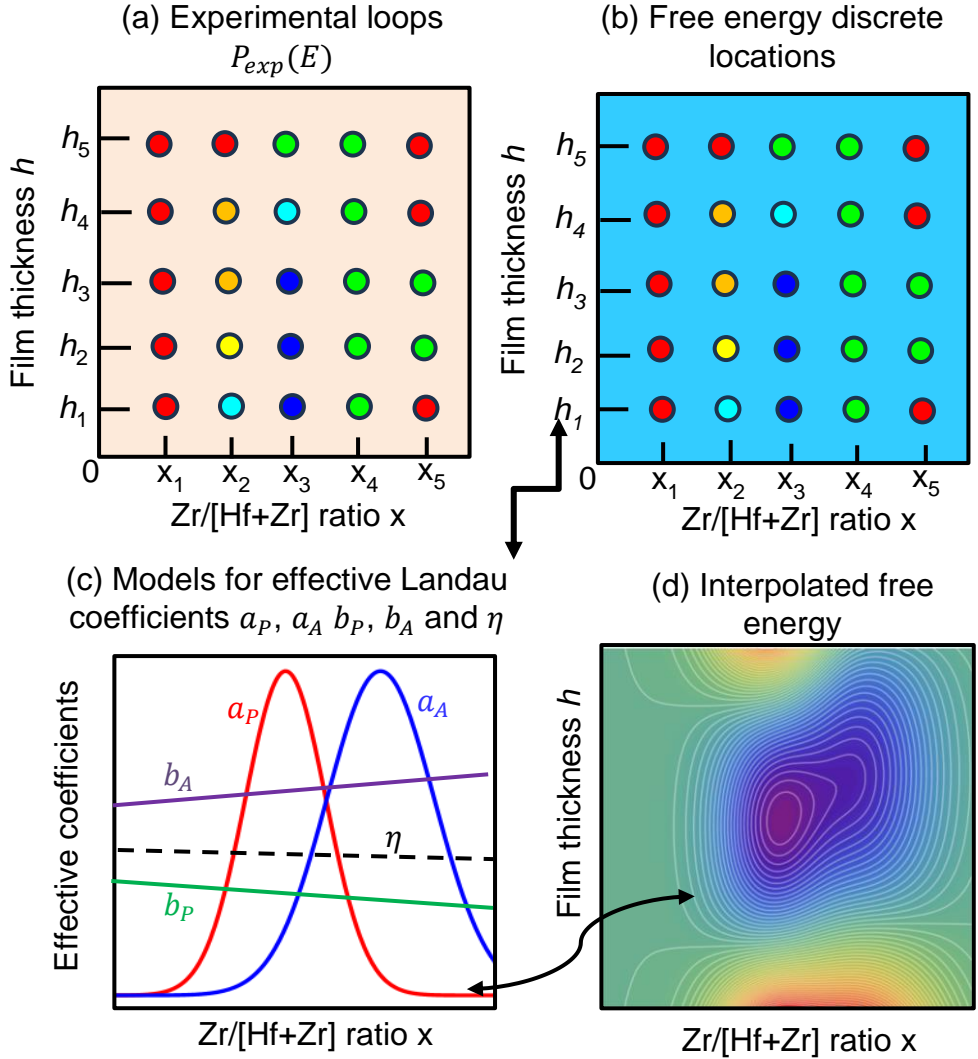

**Figure 5.** Schematic illustration for the determination of the effective Landau coefficients from the limited number experimental hysteresis loops. The part (**a**) is the experimental loops/curves for 25 pairs of $\{x_i, h_j\}$ ($i, j = 1, 2, ..., 5$),

**(b)** is the free energy sampled at the same 25 discrete locations $\{x_i, h_j\}$. The part **(c)** is the illustrative models for effective Landau coefficients $a_P$, $a_A$ $b_P$, $b_A$ and $\eta$; **(d)** is the interpolated effective Landau energy in dependence on $x$ and $h$. Red, orange, yellow, cyan, blue and green circles in the parts **(a)** and **(b)** correspond to the dielectric and paraelectric curves, paraelectric-like loops, thin ferroelectric-like loops, wide ferroelectric-type loops, and antiferroelectric-type double loops, respectively.

Since the bulk HfO$_2$ and ZrO$_2$ are neither FE nor AFE, the unusual dependence of the effective coefficients on the thickness $h$ of HZO films can be explained by the "double" size effect, which consists in the h-dependence of the average grain size $R$, being a key parameter in Eqs.(8). The nonmonotonic dependence of the grain size on the ratio $x$ is a well-known effect (see Fig. 2 in Ref. [8]).

The effective coefficients are shown in **Fig. S2** [30]. It is seen from the figure, that the smooth interpolation functions of x and $h$ exist for effective coefficients $a_P(x, h)$, $a_A(x, h)$, $b_P$, $b_A$ and $\eta$. These interpolation functions allow to describe the transformation of polarization dependences on applied electric field and hysteresis loop shapes induced by the continuous changes of x and $h$.

Noteworthy, in all our previous studies of various nanoparticles and thin films, including HfO$_2$ films, we derived or used available analytical dependences of the size- or/and thickness- dependences of the LGD expansion coefficients, because the size effect in many nanoscale ferroelectrics can be successfully described by a combination of the long-range depolarization field and strain and/or Vegard stress effects, and short-range surface and/or interface effects [33, 48, 49, 50, 51]. Disregarding our efforts, the HZO thin films appeared to be a strongly anomalous case, as argued in detail in **Appendix A** [30]. In brief, the polar properties of HZO films are conditioned by the "double" size effect (when the properties are R- and h-dependent) rather than by a conventional "single" size effect (when the properties are only h-dependent). The analytical description of the "double" size effect has not been evolved yet, and we hope that the present work can be helpful in the direction. We also note that the ab initio calculations can be very important for the correct determination of the ground state (PE, FE, FI or AFE) of an ultra-thin homogeneous epitaxial HZO film, but the films studied by Park et al. [8] have a pronounced granular structure, which makes their ab initio simulation much more complicated.

### IV. WAKE-UP EFFECT OF THE POLARIZATION REVERSAL IN THE HZO FILM

The Landau free energy (6a) of a HZO film, written via the polarization and electric field, is:

$$G_f = \frac{\alpha}{2} P_f^2 + \frac{\beta}{4} P_f^4 - E_f P_f. \qquad (12)$$

Here the coefficients $\alpha$ and $\beta$ are expressed via the coefficients $a_P$, $b_P$, $\eta$, $a_A$ and $b_A$ of the free energy (6a) in accordance with **Table I**. Namely, $\alpha = a_P$ and $\beta = b_P$ in the DE, PE, FE or AFE phases with $A_f = 0$, and $\alpha = a_P - \eta \frac{a_A}{b_A}$ and $\beta = b_P - \frac{\eta^2}{b_A}$ in the FI or AFE phases with $A_f^2 = -\frac{a_A + \eta P_f^2}{b_A}$. For β > 0 the free energy (12) allows to describe a paraelectric state with $P_f \sim E_f$ for α > 0, and two stable states with the spontaneous polarization $P_s = \pm\sqrt{-\frac{\alpha}{\beta}}$ for α < 0. Minimizing the free energy by $P_f$, we obtain the well-known expression for the polarization behavior in the electric field:

$$\alpha(x,h)P_f + \beta(x,h)P_f^3 = E_f. \quad (13)$$

In Eq.(13) the "effective" parameters $\alpha(x,h)$ and $\beta(x,h)$ depend on the annealing/deposition temperature $T_a$, Zr/[Hf+Zr] ratio $x$ and the film thickness $h$.

Let us also assume that, due to the very slow relaxation of local stresses, the change of $\alpha(x,h)$ is determined by the number $N$ of electric field cycles, and after a certain big number of cycles, $N_s$, it stops changing and reaches saturation. Formally the assumption is the most natural for LGD formalism, and from the physical viewpoint the reason may be the gradual transition of the HZO film from a series of metastable states separated by shallow energy barriers into a stable configuration separated by the highest barrier, where the system remains unchanged under subsequent cycles of repolarization.

Sketch in the top left of **Fig. 6** shows the case, when the tetragonal grains dominate, and the "wake-up" transformation of the numerous transient metastable tetragonal states to the ground orthorhombic state lasts very long and is very pronounced. Sketch in the top right of **Fig. 6** demonstrates the case, when the orthorhombic grains dominate, and the wake-up effect is very short and much less pronounced.

Note that inhomogeneous chemical stresses created by e.g., oxygen vacancies and/or defects being at the same time elastic dipoles, can contribute to the "ripples" of the multi-well potential for the polar and antipolar ordering due to the electrostriction coupling. The concentration and distribution of elastic dipoles can significantly depend on the deposition temperature of HZO layers, which may explain the difference between the wake-up characteristics of the HZO films prepared by plasma-enhanced atomic layer deposition (ALD) at deposition temperatures $T_d$ =200ºC and 250ºC in Saini et al. [28] experiments. To verify the assumption, the ground state calculations of HZO in the multi-well thermodynamic potential can be in order, to reach the relaxed state before applying the periodic electric potential. A preliminary FEM of the $P_f$ and $A_f$ relaxation in the presence of the potential "ripples" in the LGD free energy shows that the system relaxation under the absence of applied voltage depends strongly on the ripples amount and height, and also on the Landau-Khalatnikov relaxation coefficients, $\Gamma_P$ and $\Gamma_A$, in Eqs.(9). For definite relations between these parameters, which can be a question of a separate study, we observed the system "pinning" in the shallow potential wells, which leads to the multiple metastable configurations, where the system can exist during a very long time of simulations.

The simplest Pade-exponential approximation describing the relaxation mechanism yields the expressions for $\alpha$ and $P_s$:

$$\alpha(x,h,N) = \alpha_s(x,h) + [\alpha_0(x,h) - \alpha_s(x,h)]\exp\left(-\frac{N}{N_s}\right) \approx \begin{cases} \alpha_0(x,h), & N \ll N_s, \\ \alpha_s(x,h), & N \gg N_s. \end{cases} \quad (14a)$$

$$P_s(x, N, h) = \pm\sqrt{-\frac{\alpha(x,h,N)}{\beta}} \approx \begin{cases} \pm\sqrt{-\frac{\alpha_0(x,h)}{\beta}}, & N \ll N_s, \\ \pm\sqrt{-\frac{\alpha_s(x,h)}{\beta}}, & N \gg N_s. \end{cases} \quad (14b)$$

Expression (14) makes it possible to describe qualitatively the "awakening" effect of the ferroelectric phase in the thin HZO films, studied in Ref. [28]. The 10-nm thick HZO films were capped with 10-nm thick TiN top and bottom electrodes [28].

The phase composition of the HZO film, prepared by ALD at $T_d$ =200°C, contain a significant amount of metastable transient states, which are associated with the tetragonal symmetry regions (e.g., located around the Vegard-type elastic defects). During the multiple cycles of electric field, the gradual transitions from the transient states to the polar orthorhombic phase occur, and the behavior of the remanent polarization is described by expression (14b). Such an increase in polarization with an increase in the number of repolarization cycles is shown in experimental Fig. 1b in Ref.[28].

The 50°C difference in the deposition temperature can lead to the differences in the average sizes of tetragonal and orthorhombic regions, as well as facilitate the Vegard stress relaxation, which, in accordance with the model [8] and expressions (8), can make the orthorhombic phase more stable in the as-prepared films. Therefore, the phase composition of the HZO film, prepared by ALD at $T_d$ =250°C and all other similar conditions, mostly corresponds to the stable orthorhombic phase and contain only small number of metastable states. For this case the remanent polarization is virtually independent on the number of repolarization cycles (see Fig. 1c in Ref. [28]).

It should be mentioned that the 50°C difference in the deposition temperature for the first and the second HZO films does not mean that the activation barrier between the stable and metastable states is about 4 meV. The activation energy can be many times higher. Otherwise, a spontaneous return from the stable to metastable states would occur at room temperature due to the interaction with vibrational modes.

The polarization-field dependences, $P_f(E)$, calculated for $Hf_{0.5}Zr_{0.5}O_2$ films deposited on the same substrate at lower and higher temperatures are shown in **Figs. 6(a)** and **6(b)**, respectively. Polarization $P_f$ is normalized on the spontaneous value, $P_s$ =0.15 C/m², and electric field $E$ is normalized on the coercive field value, $E_c$ =0.15 V/nm. The polarization hysteresis loops are calculated from Eqs.(9) using the functions

$$a_P(N) = a_{Ps} + [a_{P0} - a_{Ps}]\exp\left(-\frac{N}{N_s}\right), \quad (15a)$$

$$a_A(N) = a_{As} + [a_{A0} - a_{As}]\exp\left(-\frac{N}{N_s}\right). \quad (15b)$$

We regard that the dimensionless parameters in Eq.(15), namely the values of $a_{P0}$ and $a_{A0}$, can depend strongly on the deposition temperature $T_d$, film thickness, and/or other preparation conditions due to several reasons. For instance, the concentration and z-distribution of oxygen vacancies and/or other elastic defects can depend strongly on the deposition conditions (see, e.g., Ref. [52]).

The conditions $a_{P0} < 0$ and $a_{A0} > 0$, used in **Fig. 6(b)**, correspond to the FE-like initial state of the HZO film with a single polarization hysteresis loop. The conditions $a_{Ps} < 0$ and $a_{As} > 0$, used in **Fig. 6(a)** and **6(b)**, correspond to the FE-like final state of the HZO film. The conditions $b_P > 0$, $b_A > 0$ and $\eta > -2\sqrt{b_P b_A}$ correspond to the functional stability in the initial and final states; the values $b_P = b_A = 1$ can be set in dimensionless units, and the value $\eta \geq 1$ agrees with the phase portrait shown in **Fig. 2(d)**. The concrete values of $a_{P0}$ and $a_{A0}$, as well as $a_{Ps}$ and $a_{As}$, were selected after their enumeration in the broad range {-5, 5}, which aim was to achieve the best agreement with the experimental results [28]. In result we obtained that the values $a_{P0} = 2$ and $a_{A0} = -2$ correspond to the deposition temperature $T_d = 200°C$ (simulation results are shown in **Fig.6(a)**); and the values $a_{P0} = -1$ and $a_{A0} = 2$ correspond to the deposition temperature $T_d = 250°C$ (simulation results are shown in **Fig.6(b)**). Other parameters $a_{Ps} = -(2.0 - 2.5)$, $a_{As} = 2$, $b_P = b_A = 1$, $\eta = 2$, and $w = 0.1$, which are selected close or the same for both values of $T_d$, are also important for the best qualitative agreement with experimental results shown in Fig. 1 in Ref.[28].

The conditions $a_{P0} = 2$ and $a_{A0} = -2$, which correspond to the high positive coefficient $a_P(N \ll N_s) \approx 2$ and the negative coefficient $a_A(N \ll N_s) \approx -2$, mean that the initial state of the HZO film deposited at 200°C is the AFE-like with double or strongly pinched polarization hysteresis loop due to the high concentration of "stubborn" elastic defects. The wake-up process, which redistribute the defects and/or eliminate their significant part, transforms the film to the FE-like state with the high negative coefficient $a_P(N \gg N_s) \approx -2$ and the positive coefficient $a_A(N \gg N_s) \approx 2$, which correspond to the single loop of polarization [see **Fig. 6(a)**].

The conditions $a_{P0} = -1$ and $a_{A0} = 2$, which correspond to the negative coefficient $a_P(N \ll N_s) \approx -1$ and the high positive coefficient $a_A(N \ll N_s) \approx 2$, mean that the initial state of the HZO film deposited at 250°C is already FE-like with saturated polarization hysteresis loop due to the low concentration of "stubborn" elastic defects. The wake-up process redistributes and/or eliminates the defects, and thus slightly improves the film FE properties, which correspond to the higher negative coefficient $a_P(N \gg N_s) \approx -2.5$ and the same positive coefficient $a_A(N \gg N_s) \approx 2$ [see **Fig. 6(b)**].

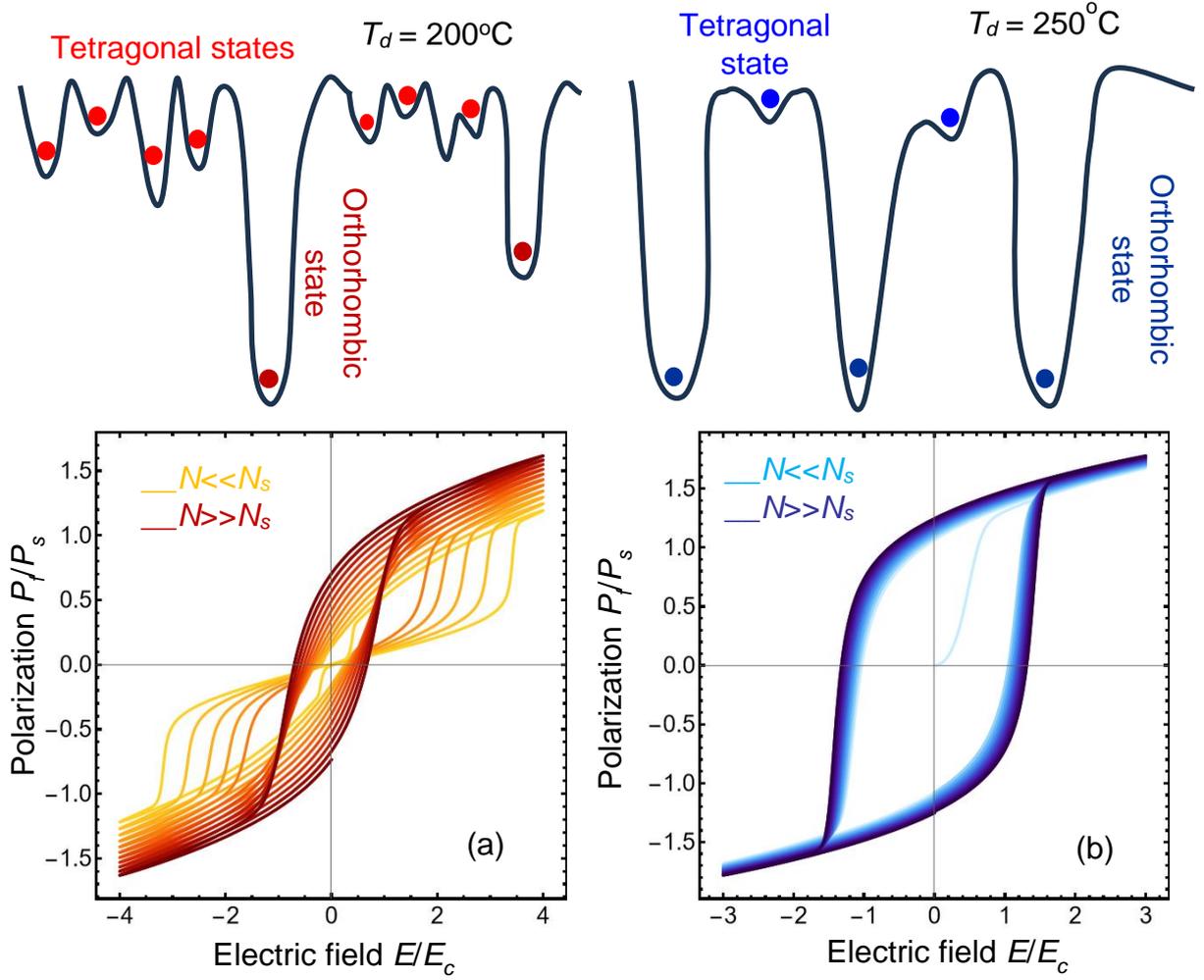

**Figure 6.** The polarization-field dependences, $P_f(E)$, calculated for a 10-nm Hf$_{0.5}$Zr$_{0.5}$O$_2$ film deposited on the same substrate at lower and higher deposition temperatures, $T_d = 200$ºC **(a)** and 250ºC **(b)**, respectively. Polarization $P_f$ is normalized on the spontaneous value, $P_s \cong 0.15$ C/m$^2$, and electric field $E$ is normalized on the coercive field value, $E_c = 0.15$ V/nm. The polarization loops are calculated from Eqs.(9). Dimensionless parameters $a_{P0} = 2$, $a_{Ps} = -2.0$, $a_{A0} = -2$, and $a_{As} = 2$ for the plot **(a)**, $a_{P0} = -1$, $a_{Ps} = -2.5$, $a_{A0} = 2$, and $a_{As} = 2$ for the plot **(b)**, $b_P = b_A = 1$, $\eta = 2$, and $w = 0.1$ are selected to reach the best agreement with experimental results [28]. The color changes from the light yellow (for the first loops, $N \ll N_s$) to the dark red (for the last loops, $N \gg N_s$) in the part **(a)**, and from the light cyan (for the first loops, $N \ll N_s$) to the dark blue (for the last loops, $N \gg N_s$) in the part **(b)**.

### V. THE INFLUENCE OF HZO POLARIZATION ON GRAPHENE CONDUCTIVITY

Since a single-layer graphene is a Dirac semiconductor with very high concentration of carriers, it effectively screens the electric polarization of the HZO film and prevents the domain formation in the FE phase. At the same time, the polarization influences strongly on the graphene conductivity, and the conductivity can be estimated according to the simple Drude formula [53]

$$\sigma_g \cong e n_{2D} \mu, \quad (16)$$

where $e = 1.602 \cdot 10^{-19}$ C is the elementary charge, $n_{2D}$ is the 2D-concentration of free carriers in the single-layer graphene, and $\mu$ is their mobility. For the most often common case, when the scattering in the graphene channel mainly occurs on ionized impurities outside it, $\mu$ is almost constant and can be estimated as $(0.5 - 10)$ m²/Vs [53]. The magnitude of $n_{2D}$ is related with the 3D-concentration, $n$, as $n_{2D}(x, y) = \int_0^{h_g} n(x, y, z) dz$.

Assuming that the dielectric gap between the graphene layer and the HZO film is absent, and the boundary condition $D_3^g(x, y, -h_g) = 0$ is valid at the graphene surface, the magnitude of $n_{2D}$ is determined by the film polarization $P_f$, which is regarded single-domain and quasi-homogeneous. For the case $n_{2D}$ can be estimated as [16, 17]:

$$n_{2D} \approx \frac{1}{e} |P_f + \varepsilon_0 E_3|. \quad (17)$$

Since the soft mode is absent in a bulk HZO, we regard that $\varepsilon_{ij}^b \approx \delta_{ij}$ and thus $D_3^g = P_f + \varepsilon_0 E_3$. The field $E_3 \approx \frac{V}{h+h_g}$ in Eq.(17). Let us underline that the accuracy of approximate equality in Eq.(17) becomes low for a strongly inhomogeneous distribution of polarization created by e.g., a 180-degree domain structure with a period smaller than the film thickness $h$. For the case $|P_f + \varepsilon_0 E_3|$ in the right-hand side of Eq.(17) should be substituted by an unknown value averaged over the domain depth. Alternatively, the boundary problem formulated by Eqs.(1)-(9) should be solved numerically.

The field-dependence of the carrier concentration in a single-layer graphene, $n_{2D}(E)$, which is in a perfect electric contact with a Hf$_{1-x}$Zr$_x$O$_2$ film, is shown in **Fig. 7**. The concentration $n_{2D}$ increases from $10^{17}$ m$^{-2}$ to $10^{18}$ m$^{-2}$ with $E$ increase. The electric field $E$ is normalized on the HZO coercive field value, $E_c = 0.15$ V/nm; the graphene thickness $h_g = 0.4$ nm, and dimensionless LGD-parameters are the same as in **Fig. 3.** Since the thickness of a graphene single-layer measured by AFM varies in the range $(0.4 - 1.7)$ nm depending on the structure of the ripples [54], and the mobility changes in the range $(0.5 - 10)$ m²/Vs [53], the carrier concentration can vary from $10^{17}$ m$^{-2}$ to $10^{18}$ m$^{-2}$, respectively. We checked numerically, that, since the concentrations in the order of $10^{17}$ m$^{-2}$ or more are sufficiently high, they provide an almost perfect screening of the ferroelectric polarization in the HZO film and thus the formation of 180-degree domains is energetically unfavorable [16, 40]. Actually, the domain formation occurs in the case of imperfect screening (e.g., when the dielectric gap between the graphene and the HZO film exists) if the positive energy of domain walls is smaller than the decrease in the electrostatic energy related with the domain formation [40]. Hence, the assumption about the single-domain state of the HZO film is self-consistent and the approximate expression (17) is valid with high accuracy for the quasi-homogeneous single-domain polarization.

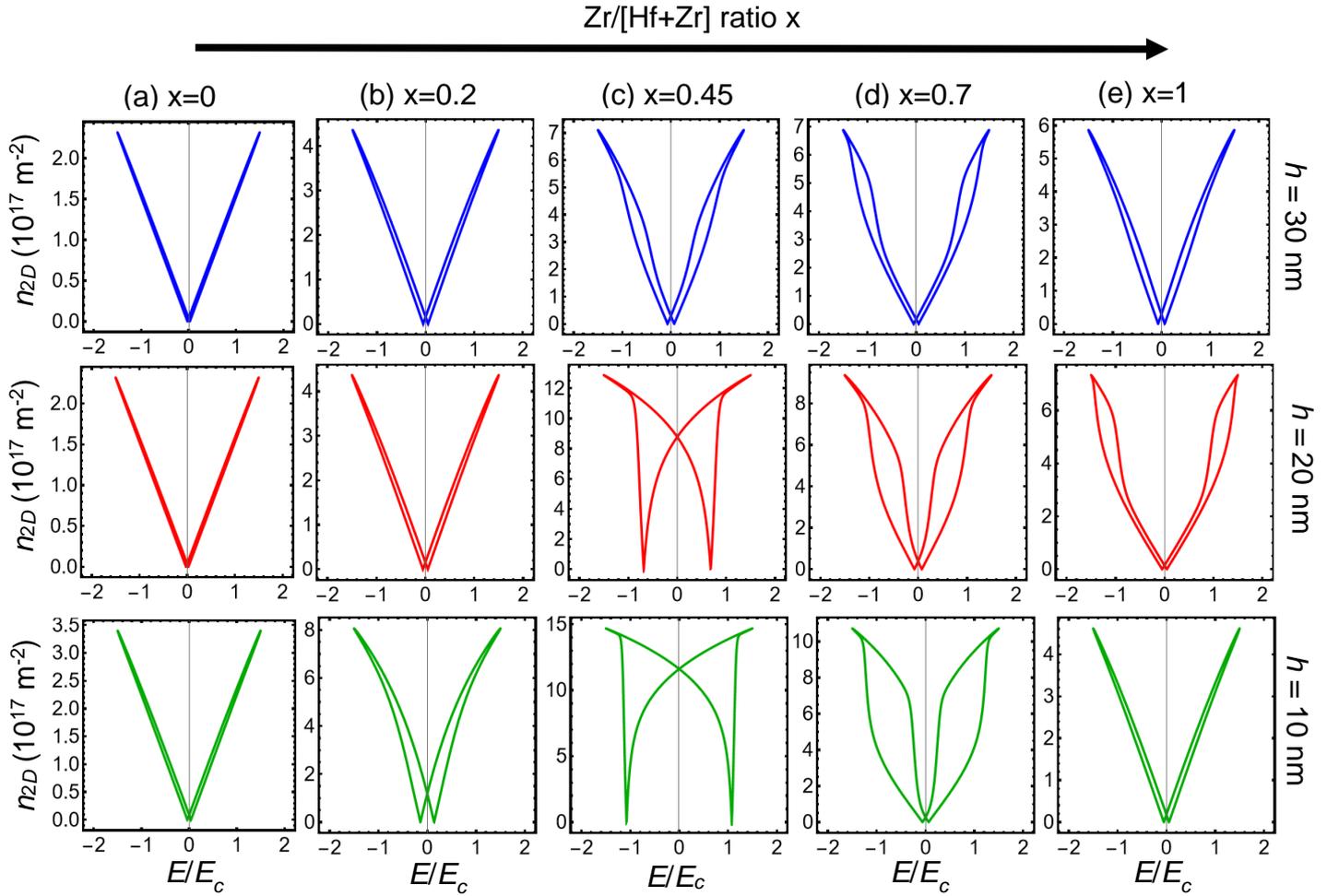

**Figure 7.** The field dependence of the 2D carrier concentration, $n_{2D}(E)$, in the single-layer graphene, which is in a perfect electric contact with a Hf$_{1-x}$Zr$_x$O$_2$ film. The ratio $x$ = 0.00 **(a)**, 0.20 **(b)**, 0.45 **(c)**, 0.70 **(d)**, 1.00 **(e)**, film thickness $h$ = 10 nm (the top row), 20 nm (the middle row) and 30 nm (the bottom row). The electric field $E$ is normalized on the HZO coercive field value, $E_c$ = 0.15 V/nm; $h_g$ =0.4 nm, and dimensionless LGD-parameters are the same as in **Fig. 3.**

Notice, that the polarization of the thinnest 10-nm HZO film induces the V-like hysteresis-less dependence $n_{2D}(E)$ for x ≤ 0.3; the "wide" double hysteresis of $n_{2D}(E)$ appears for 0.35 ≤ x ≤ 0.55, the "narrow" double hysteresis loop of $n(E)$ occurs for 0.6 ≤ x ≤ 0.9, and the $n(E)$ becomes almost hysteresis-less again for x > 0.90. The polarization of the 20-nm HZO film induces the V-like hysteresis-less dependence $n_{2D}(E)$ for x ≤ 0.35, the wide double hysteresis of $n_{2D}(E)$ appears for 0.4 ≤ x ≤ 0.55, and the loop significantly narrows for 0.7 ≤ x ≤ 1. The polarization of the thickest 30-nm film induces the V-like hysteresis-less dependence $n_{2D}(E)$ for x ≤ 0.4, the thin double hysteresis loop of $n_{2D}(E)$ appears for 0.45 ≤ x ≤ 0.95, and then $n_{2D}(E)$ becomes almost hysteresis-less for x > 0.95.

From the comparison of **Fig. 7** and **Fig. 3**, the full correlation between the HZO polarization and charge carrier concentration in graphene is evident, namely $n_{2D} \sim |P_f|$, which is in a complete agreement with the approximate expression (17). Thus, in accordance with our modeling, the polarization of the HZO

films of thickness (5 – 25) nm and Zr ratio $0.35 \leq x \leq 0.95$, which are in the FE, FI or AFE state, determine the concentration of carriers in graphene and can control its field dependence. In principle, the result can be promising for creation of concurrent Si-compatible nonvolatile memories and next generation of graphene-ferroelectric FETs, if the working voltage applied to the HZO film (which acts as a gate) is not very high. For the considered case the coercive voltage, $V_c = hE_c$, non-monotonically changes from 0.3 V to 1.6 V with $x$ increase for the 10-nm HZO film-gate (see the bottom row in **Fig. 7**); from 0.4 V to 2.1 V with $x$ increase for the 20-nm HZO film-gate (see the middle row in **Fig. 7**); and does not exceed 4.5 V for the 30-nm HZO film-gate (see the top row in **Fig. 7**). These relatively low voltages are sufficient to induce the hysteresis of graphene charge and ferroelectric polarization in the HZO gate. However, the coercive voltages are calculated for dimensionless parameters listed in **Table II**, which provide a qualitative agreement with experimental results [8]. For quantitative agreement we should use the LGD parameters in physical units determined in **Appendix A** [30]**.**

The voltage dependence of the conductivity in a single-layer graphene, $\sigma_g(V)$, are shown in **Fig. 8** for HZO films with thickness around 10 nm, $x$ =0.45 and 0.7. Here we use the LGD parameters of HZO from **Table SI** [30]. From **Figs. 8(a)** and **8(c)** the coercive voltage for SLs is (1.5 – 2) V; and the first critical voltage for DLs, shown in **Figs. 8(b)** and **8(d)**, is (0.75 – 2.05) V, and the second critical voltage can be less than (3 – 4) V. The magnitude of the remanent HZO polarization is about (0.15 – 0.2) C/m$^2$ and the graphene conductivity increases up to 0.25 S/□ under the voltage increase. Therefore, we can conclude that relatively low gate voltages (less than 2 V) applied to the HZO gate are sufficient to induce the pronounced hysteresis of the graphene conductivity.

**Figs. 8(c)** and **8(d)** are built in the approximation that all carriers in graphene are created only by the ferroelectric gate electric field. However, graphene also has its own carriers caused by thermal transitions between the valence and the conduction bands, the concentration of which at $T = 300$ K is of the order of $10^{15}$ m$^{-2}$ [55]. However, as can be seen from the carrier concentrations shown in **Fig. 7**, this effect is important only in the vicinity of the Dirac point, where it leads to a small but nonzero value of the specific conductivity.

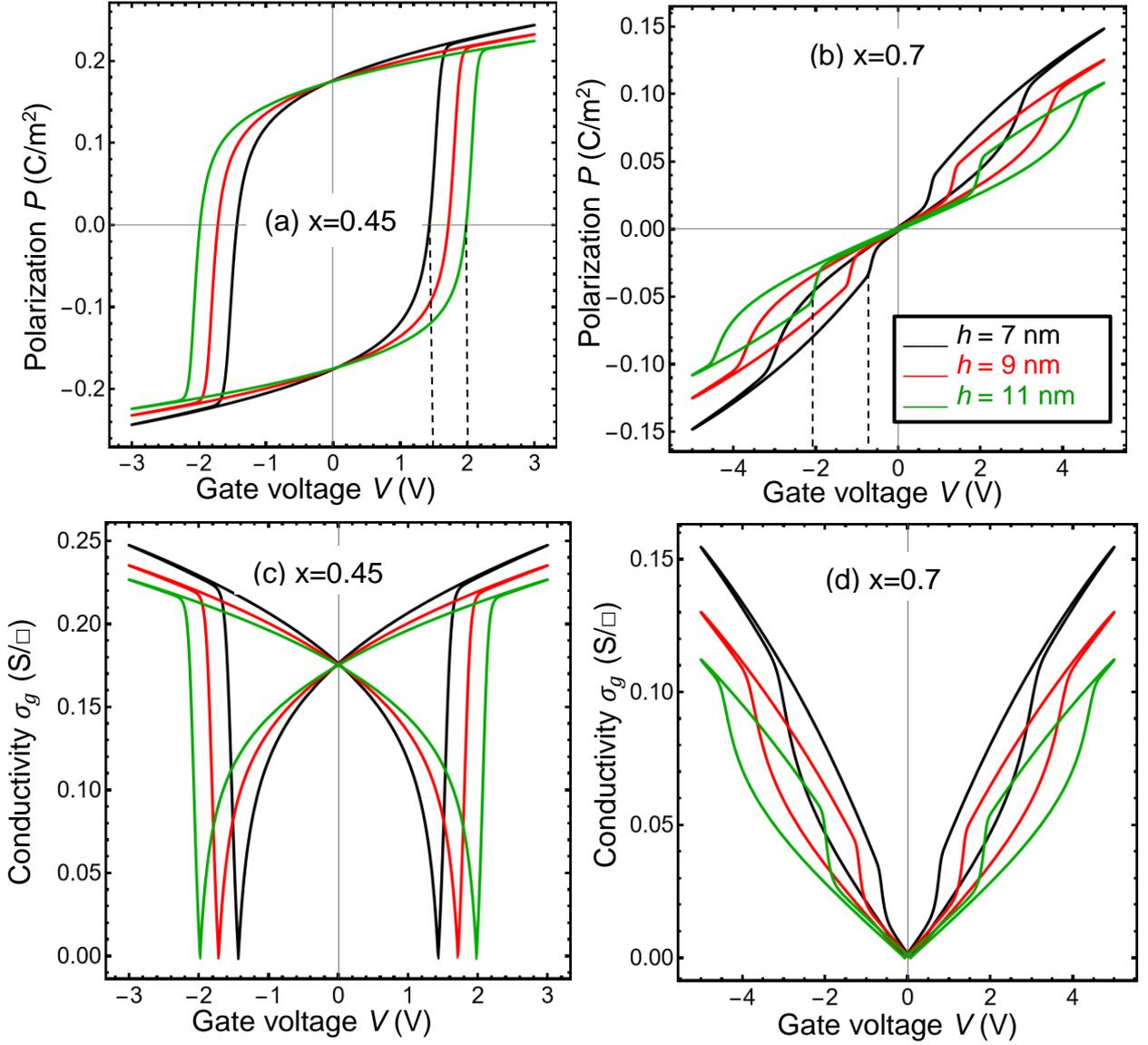

**Figure 8.** Voltage dependences of the polarization $P(V)$ **(a, b)** and conductivity of the single-layer graphene, $\sigma_g(V)$, **(c, d)**, which is in a perfect electric contact with the $Hf_{1-x}Zr_xO_2$ film of thickness $h = 7$ nm (black curves), 9 nm (red curves) or 11 nm (green curves). The Zr/[Hf+Zr] ratio $x = 0.45$ **(a, c)** and 0.70 **(b, d)**; $\mu = 1$ m$^2$/Vs, $h_g = 0.4$ nm, and LGD-parameters are given in Appendix A [30].

Finally, let us discuss the situation, when the single-layer graphene is separated from the HZO film by an ultra-thin dielectric gap (e.g., less than 1 nm and more than 0.1 nm), and the graphene acts as the top grounded electrode. In the case the domain stripes occur in the HZO film because their formation leads to the decrease of the system electrostatic energy [56]. Due to the stray field in the gap, the p-n junctions in graphene layer are induced by the domain stripes in the HZO film, at that it is important to control the number of the domain walls, each of which induces the p-n junction [16 - 26]. The motion and reversal of the domains by the dragging electric field, applied along the graphene channel, can be modeled using the self-consistent LGD approach combined with classical electrostatics [see Eqs.(1)-(9)]. It appears that relatively low periodic voltages (~ 0.5 - 2 V) can induce the robust hysteresis of the HZO polarization and graphene charge. In accordance with our calculations, the predicted effects become pronounced if the HZO film is thin enough (5 nm < $h$ < 20 nm), Zr/[Hf+Zr] ratio $0.3 < x < 0.7$, and the gap between the film and the graphene layer is ultra-thin (< 0.5 nm). Under the validity of these conditions, the nanostructure single-layer graphene - dielectric gap - HZO film-gate can be a promising candidate for the Si-compatible graphene-ferroelectric FETs, modulators and rectifiers based on the graphene p-n junctions.

## VI. CONCLUSIONS

Using LGD approach, we predicted the strong charge-polarization coupling in the nanostructure "thin HZO film – single-layer graphene" and calculated the polarization of the HZO film, which are strongly affected by the Zr/[Hf+Zr] ratio x, grain size and thickness effects. We also evolved a simple phenomenological model for the description of the polarization wake-up in the HZO films and its dependence on deposition conditions.

Using a limited number of polarization-field curves and hysteresis loops measured for thin HZO films capped with conducting TiN electrodes, assuming full screening of the polarization in the electrodes, and the absence of surface electrochemical effects, we determine effective expansion coefficients of the Landau free energy. The effective coefficients are nonlinearly dependent on the film thickness $h$ and Zr/[Hf+Zr] ratio x, in contrast to $h$-independent and linearly x-dependent expansion coefficients of a classical Landau energy describing FE-AFE solid solutions, such as a bulk $Pb_{1-x}Zr_xO_3$. Since the bulk $HfO_2$ and $ZrO_2$ are neither FE nor AFE, we explain the unusual dependence of the Landau expansion coefficients by the "double" size effect. The effect consists in the anomalous nonmonotonic dependence of the average grain size $R$ and surface energy on the thickness $h$ of the HZO film, which is conditioned by the available experimental data for 9 nm < $h$ < 29 nm [8]. The Landau free energy with five "effective" expansion coefficients, which are interpolation functions of x and $h$, describes the transformation of polarization dependences on applied electric field and hysteresis loop shapes induced by the continuous changes of x and $h$.

Using the effective Landau free energy, we revealed the full correlation between the distribution of polarization in the HZO film and charge carriers in the single-layer graphene. In particular, the polarization of the (5 – 25) nm thick HZO films, which are in the ferroelectric-like or antiferroelectric-like states for the range $0.35 \leq x \leq 0.95$, determine the concentration of carriers in graphene and can control its field dependence. The result can be promising for creation of next generation Si-compatible FeRAMs and FETs, because the working voltage applied to the HZO film (which acts as a gate) is less than (1.5 – 2) V. These relatively low voltages are sufficient to induce the pronounced hysteresis of ferroelectric polarization in the HZO gate, which induces the hysteresis of the graphene charge and conductivity.

When the single-layer graphene is separated from the HZO film by an ultra-thin dielectric gap, the domain stripes occur in the film. Simultaneously, the p-n junctions in graphene layer are induced by the domain stripes. Under favorable conditions, the nanostructure "single-layer graphene - dielectric gap - HZO film-gate" can be a promising candidate for the Si-compatible modulators and rectifiers based on the p-n junctions induced in the graphene.

**Data availability statement.** Numerical results presented in the work are obtained and visualized using a specialized software, Mathematica 13.2 [57]. The Mathematica notebooks, which contain the codes, are available per reasonable request (https://notebookarchive.org/2023-07-7ffe0qr).

**Authors' contribution.** The research idea belongs to A.N.M., M.V.S. and S.V.K. A.N.M. and M.V.S. formulated the problem, performed analytical calculations, analyzed results, and wrote the manuscript draft. E.A.E. wrote codes. S.V.K. worked on the results explanation and manuscript improvement. All co-authors discussed the obtained results.

**Acknowledgments.** Authors are very grateful to the Referees for constructive suggestions and useful discussions. This work was partially supported by the center for 3D Ferroelectric Microelectronics (3DFeM), an Energy Frontier Research Center funded by the U.S. Department of Energy (DOE), Office of Science, Basic Energy Sciences under Award Number DE-SC0021118. This work was partially supported by the U.S. Department of Energy, Office of Science User Facility. A.N.M. work is supported by the Ministry of Science and Education of Ukraine (grant № PH/ 23 - 2023, "Influence of size effects on the electrophysical properties of graphene-ferroelectric nanostructures") at the expense of the external aid instrument of the European Union for the fulfillment of Ukraine's obligations in the Framework Program of the European Union for scientific research and innovation "Horizon 2020". A.N.M. is very grateful to Dr. Yuri Zagorodny for useful discussions.

Anna N. Morozovska[1], Maksym V. Strikha [2, 3*], Kyle P. Kelley[4], Sergei V. Kalinin[5†],

and Eugene A. Eliseev[6‡]

[1] Institute of Physics, National Academy of Sciences of Ukraine,

Pr. Nauky 46, 03028 Kyiv, Ukraine,

[2] Taras Shevchenko National University of Kyiv, Faculty of Radiophysics, Electronics and

Computer Systems, Pr. Akademika Hlushkova 4g, 03022 Kyiv, Ukraine,

[3] V. Lashkariov Institute of Semiconductor Physics, National Academy of Sciences of Ukraine,

Pr. Nauky 41, 03028 Kyiv, Ukraine

[4] Center for Nanophase Materials Sciences, Oak Ridge National Laboratory,

Oak Ridge, TN 37831, USA

[5] Department of Materials Science and Engineering, University of Tennessee,

Knoxville, TN, 37996, USA

[6] Institute for Problems of Materials Science, National Academy of Sciences of Ukraine,

Krjijanovskogo 3, 03142 Kyiv, Ukraine

## APPENDIX A. Determination of the effective Landau expansion coefficients from the experimental results

Using the coupled system of static equations (9), written as following:

$$a_P P_f + b_P P_f^3 + \eta A_f^2 P_f = E_3, \qquad (A.1a)$$

$$a_A A_f + b_A A_f^3 + \eta P_f^2 A_f = 0, \qquad (A.1b)$$

it is possible to derive analytical expressions for the order parameters, their features and critical fields, which are listed in the first and last columns of **Table I.** Using the relaxation-type equation

$$\Gamma_P \frac{\partial P_f}{\partial t} + \alpha P_f + \beta P_f^3 = E_3, \qquad (A.2)$$

we can try to fit the shape of the loops, shown in Fig. 1 in Ref. [1]. Since the coercive field depends on several factors, not included in the model (A.2), it seems more reasonable to fit the values of remanent polarization, $P_r$, and its slope, which is the dielectric susceptibility, $\chi = \frac{dP_r}{dE_3}\Big|_{E_3 \to 0}$, in the small quasi-static electric fields. The term "quasi-static" means that we assume that the frequency $\omega$ of external field $E_3$ is much smaller than the value $\frac{2\pi}{\Gamma_P}|\alpha|$. This gives the system of two coupled algebraic equations for $\alpha$ and $\beta$ determination from the static curves:

$$\alpha P_r + \beta P_r^3 = 0, \tag{A.3a}$$

$$\alpha + 3\beta P_r^2 = \frac{1}{\chi}. \tag{A.3b}$$

* Corresponding author: maksym.strikha@gmail.com

† Corresponding author, e-mail: sergei2@utk.edu

‡ Corresponding author: eugene.a.eliseev@gmail.com

Note that the relative dielectric susceptibility, $\tilde{\chi}$, measured independently, should be close to those determined from the quasi-static polarization hysteresis loops. They should coincide in the ideal case, $\chi = \varepsilon_0 \tilde{\chi}$. Thus, from Eq.(A.3):

$$\alpha = -\frac{1}{2\chi}, \quad \beta = \frac{1}{\chi P_r^2}, \quad \text{when} \quad P_r \neq 0, \tag{A.4a}$$

$$\alpha = \frac{1}{\chi}, \quad \text{when} \quad P_r = 0. \tag{A.4b}$$

Using the expressions (A.4) and guided by the dynamic and static curves in **Fig. 2**, we determined the parameters $\alpha$ and $\beta$ from $P_r$ and $\tilde{\chi}$ measured by Park et al. [1], which are listed in the first columns of **Table SI**. The parameters $\alpha$ and $\beta$ are listed in the middle columns of **Table SI**. Noteworthy that very thin, slim and/or banana-shaped [2] dynamic polarization loops, shown in Fig. 1 in Ref. [1], correspond to DC and PC static curves, respectively, and so the determination of the coefficient $\beta$ is not possible for the loops. The DE and/or PE character of the loops is confirmed by the hysteresis-less field behavior of the dielectric permittivity shown in Fig. S1 in Ref. [1].

Note that the coefficients $\alpha = a_P$ and $\beta = b_P$ in the states with $A_f = 0$; and $\alpha = a_P - \eta \frac{a_A}{b_A}$ and $\beta = b_P - \frac{\eta^2}{b_A}$ in the states with nonzero $A_f^2 = -\frac{a_A + \eta P_f^2}{b_A}$ which exist for $P_f^2 < -\frac{a_A}{\eta}$. The critical values of polarization, which correspond to the critical fields of the AFE loop opening, are equal to $P_{c1}^2 = -\frac{a_A}{\eta}$ and $P_{c2}^2 = \frac{a_P b_A - \eta a_A}{3(\eta^2 - b_P b_A)}$, respectively (see **Fig. 3** in the main text). The critical values are absent for the DCs and PCs; they transform to $P_r$ values for SLs. Thus, LGD-model parameters $a_P, a_A, b_P, b_A$ and $\eta$ satisfy the system of four nonlinear algebraic equations:

$$a_P = \begin{cases} \alpha & \text{for the PE, AFE and FE states with } A_f = 0, \\ \alpha + \eta \frac{a_A}{b_A} & \text{for the FI and AFE states with } A_f \neq 0, \end{cases} \tag{A.5a}$$

$$b_P = \begin{cases} \beta & \text{for the PE, AFE and FE states with } A_f = 0, \\ \beta + \frac{\eta^2}{b_A} & \text{for the FI and AFE states with } A_f \neq 0, \end{cases} \tag{A.5b}$$

$$\frac{a_A}{\eta} = -P_{c1}^2 \quad \text{for the AFE phase and } \eta > \sqrt{b_P b_A} > 0, \tag{A.5c}$$

$$\frac{a_P b_A - \eta a_A}{3(\eta^2 - b_P b_A)} = P_{c2}^2 \quad \text{for the AFE phase and } \eta > \sqrt{b_P b_A} > 0. \tag{A.5d}$$

Elementary transformations of Eqs.(A.5) lead to the solution for the AFE loops:

$$a_A = -\eta P_{c1}^2, \quad a_P = \alpha - \frac{\eta^2}{b_A} P_{c1}^2, \quad \eta^2 = b_A \left( b_P + \frac{\alpha}{3 P_{c2}^2} \right), \quad \text{valid for} \quad \alpha \geq 0. \tag{A.6}$$

Thus, using the solution (A.6) and limiting our consideration by the case $b_P = b_A = b$, which is in a reasonable agreement with **Fig. 3** and experimental loop shapes [1], Eqs.(A.5) can be rewritten as following:

$$b_P = b_A = \beta, \tag{A.7a}$$

$$\eta = \begin{cases} \sqrt{\beta\left(\beta + \frac{\alpha}{3P_{c2}^2}\right)} & \text{in the AFE states}, \\ \text{impossible to define in the PE and FE states}, \end{cases} \tag{A.7b}$$

$$a_A = \begin{cases} -\eta P_{c1}^2 & \text{in the AFE states}, \\ \text{impossible to define in the PE and FE states}, \end{cases} \tag{A.7c}$$

$$a_P = \begin{cases} \alpha & \text{in the DE, PE and FE states}, \\ \alpha - \frac{\eta^2}{\beta} P_{c1}^2 & \text{in the FI and AFE states}. \end{cases} \tag{A.7d}$$

The LGD-model parameters, determined from the Eqs.(A.7) and experimental results [1], are listed in the last four columns of **Table SI.**

**Table SI.** Landau-Ginsburg-Devonshire parameters of $Hf_{1-x}Zr_xO_2$ films

| $x$ | $P_r(x)$ C/m$^2$ | $P_{c1,2}(x)$ C/m$^2$ | $\tilde{\chi}(x)$ | $\chi(x)$ ×10$^{-10}$ F/m | $\alpha(x)$ ×10$^9$ m/F | $\beta(x)$ ×10$^{10}$ Vm$^5$/C$^3$ | $a_P(x)$ ×10$^9$ m/F | $a_A(x)$ ×10$^9$ m/F | $b_P \approx b_A$ ×10$^{10}$ Vm$^5$/C$^3$ | $\eta(x)$ ×10$^{10}$ Vm$^5$/C$^3$ |
|---|---|---|---|---|---|---|---|---|---|---|
| | | | | | Film thickness $h = 9.2$ nm | | | | | |
| 0 | 0 | A | 20 | 1.77 | 5.647 | N/P | $\alpha$ | N/P | N/P | N/P |
| 0.19 | 0.025 | A | 22 | 1.95 | 5.134 | N/P | $\alpha$ | N/P | N/P | N/P |
| 0.43 | 0.175 | A | 32.5 | 2.88 | -1.738 | 6.012 | $\alpha$ | N/P | $\beta$ | N/P |
| 0.70 | -0.025 | 0.05 0.175 | 42.5 | 3.76 | 2.657 | 6.012 | $\alpha$ $\alpha - \frac{\eta^2 P_{c1}^2}{\beta}$ | $-\eta P_{c1}^2$ | $\beta$ | $\sqrt{\beta\left(\beta + \frac{\alpha}{3P_{c2}^2}\right)}.$ |
| 1.00 | 0 | 0 | 35 | 3.10 | 3.227 | N/P | $\alpha$ | N/P | N/P | N/P |
| | | | | | Film thickness $h = 14.2$ nm | | | | | |
| 0 | 0 | A | 19.5 | 1.73 | 5.792 | N/P | $\alpha$ | N/P | N/P | N/P |
| 0.19 | 0.0125 | A | 22 | 1.95 | 5.134 | N/P | $\alpha$ | N/P | N/P | N/P |
| 0.43 | 0.15 | A | 32.5 | 2.88 | -1.738 | 7.722 | $\alpha$ | N/P | $\beta$ | N/P |
| 0.70 | 0.048 | 0.04 0.15 | 40 | 3.54 | 2.824 | 7.722 | $\alpha$ $\alpha - \frac{\eta^2 P_{c1}^2}{\beta}$ | $-\eta P_{c1}^2$ | $\beta$ | $\sqrt{\beta\left(\beta + \frac{\alpha}{3P_{c2}^2}\right)}.$ |
| 1.00 | -0.025 | 0.09 0.175 | 37.5 | 3.32 | 3.012 | 7.722 | $\alpha - \frac{\eta^2 P_{c1}^2}{\beta}$ | $-\eta P_{c1}^2$ | $\beta$ | $\sqrt{\beta\left(\beta + \frac{\alpha}{3P_{c2}^2}\right)}..$ |
| | | | | | Film thickness $h = 19.2$ nm | | | | | |
| 0 | 0 | A | 18 | 1.59 | 6.274 | N/P | $\alpha$ | N/P | N/P | N/P |
| 0.19 | 0.01 | A | 21 | 1.86 | 5.378 | N/P | $\alpha$ | N/P | N/P | N/P |
| 0.43 | 0.15 | A | 33 | 2.92 | -1.711 | 7.605 | $\alpha$ | N/P | $\beta$ | N/P |
| 0.70 | 0.04 | 0.07 0.15 | 32.5 | 2.88 | 3.475 | 7.605 | $\alpha$ $\alpha - \frac{\eta^2 P_{c1}^2}{\beta}$ | $-\eta P_{c1}^2$ | $\beta$ | $\sqrt{\beta\left(\beta + \frac{\alpha}{3P_{c2}^2}\right)}.$ |
| 1.00 | 0.015 | 0.08 0.125 | 35 | 3.10 | 3.227 | 7.605 | $\alpha - \frac{\eta^2 P_{c1}^2}{\beta}$ | $-\eta P_{c1}^2$ | $\beta$ | $\sqrt{\beta\left(\beta + \frac{\alpha}{3P_{c2}^2}\right)}..$ |
| | | | | | Film thickness $h = 24.2$ nm | | | | | |
| 0 | 0 | A | 20 | 1.77 | 5.647 | N/P | $\alpha$ | N/P | N/P | N/P |
| 0.19 | 0.005 | A | 18.5 | 1.64 | 6.105 | N/P | $\alpha$ | N/P | N/P | N/P |
| 0.43 | 0.051 | A | 26.5 | 2.35 | 4.262 | N/P | $\alpha$ | N/P | N/P | N/P |
| 0.70 | 0.01 | 0.065 0.10 | 28 | 2.48 | 4.034 | 7.605 | $\alpha$ $\alpha - \frac{\eta^2 P_{c1}^2}{\beta}$ | $-\eta P_{c1}^2$ | $\beta$ | $\sqrt{\beta\left(\beta + \frac{\alpha}{3P_{c2}^2}\right)}..$ |
| 1.00 | 0 | A | 30 | 2.66 | 3.765 | N/P | α | N/P | N/P | N/P |
| Film thickness $h = 29.2$ nm | | | | | | | | | | |
| 0 | 0 | A | 18.5 | 1.64 | 6.105 | N/P | α | N/P | N/P | N/P |
| 0.19 | 0 | A | 20.5 | 1.81 | 5.509 | N/P | α | N/P | N/P | N/P |
| 0.43 | 0.014 | 0.055 0.090 | 25 | 2.21 | 4.518 | 7.605 | α $\alpha - \frac{\eta^2 P_{c1}^2}{\beta}$ | $-\eta P_{c1}^2$ | β | $\sqrt{\beta \left(\beta + \frac{\alpha}{3 P_{c2}^2}\right)}..$ |
| 0.70 | 0.018 | 0.050 0.115 | 30 | 2.66 | 3.765 | 7.605 | α $\alpha - \frac{\eta^2 P_{c1}^2}{\beta}$ | $-\eta P_{c1}^2$ | β | $\sqrt{\beta \left(\beta + \frac{\alpha}{3 P_{c2}^2}\right)}..$ |
| 1.00 | 0 | A | 29 | 2.57 | 3.895 | N/P | α | N/P | N/P | N/P |

* "A" means "absent", "N/P" means "not possible to determine from the experimental data".

The coefficients $\alpha(x, h)$ and $\beta(x, h)$, determined from Fig. 1 and S1 in Ref. [1] by using Eqs.(A.4), are shown in **Fig. S1**. Symbols correspond to the values, determined from Ref. [1], solid curves are the second order interpolation functions.

The coefficients $a_P(x, h)$, $a_A(x, h)$, $b_{P,A}(x, h)$, and $\eta(x, h)$ determined from Eq.(A.7) and **Table SI**, are shown in **Fig. S2.** Symbols correspond to the points extracted from Ref. [1], solid curves are the interpolation functions.

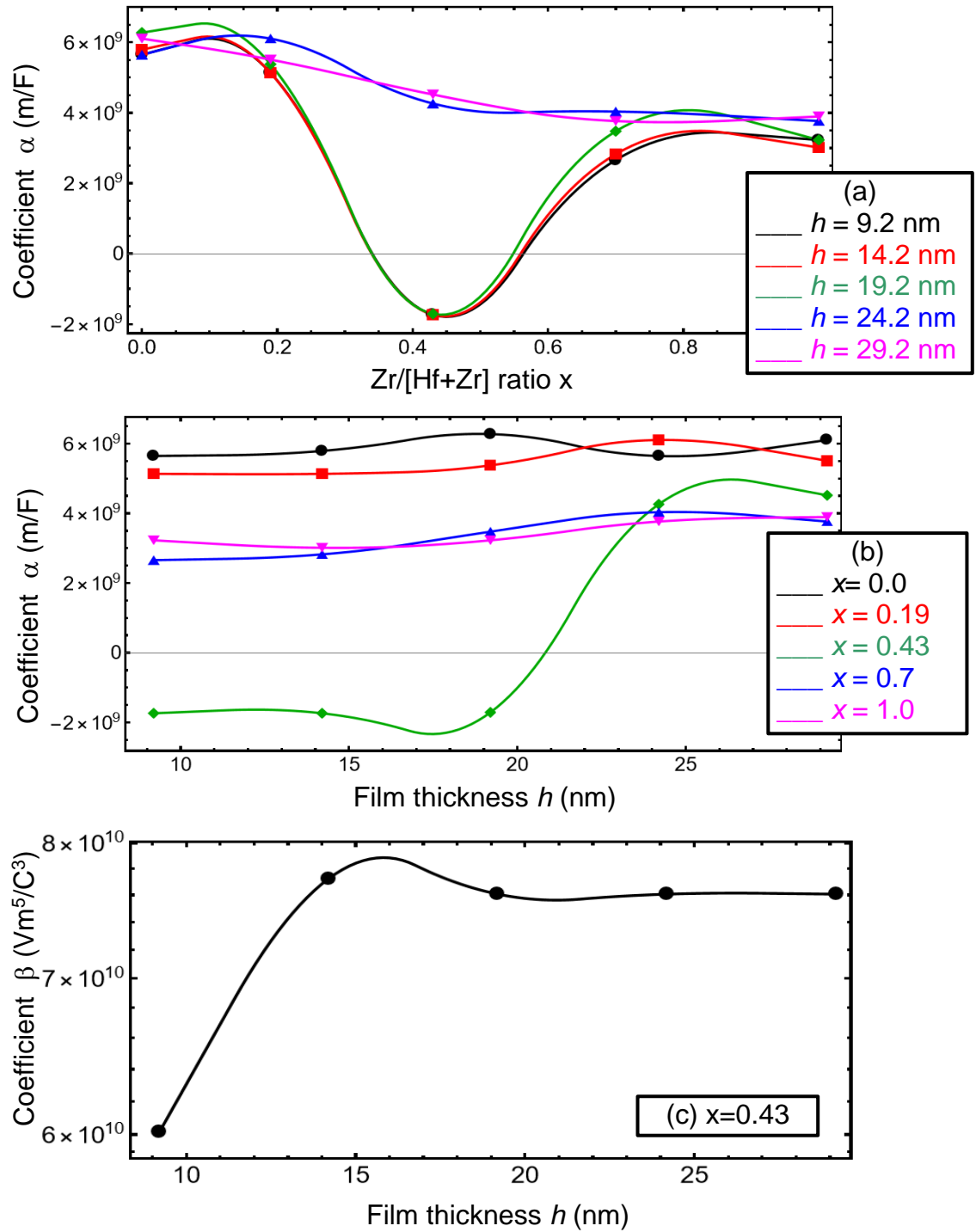

**Figure S1**. The coefficients α(x, h) **(a, b)** and β(x, h) **(c)** determined from Eqs.(A.4) and Ref. [1]. Symbols correspond to the values, determined from Ref. [1], solid curves are the second order interpolation functions.

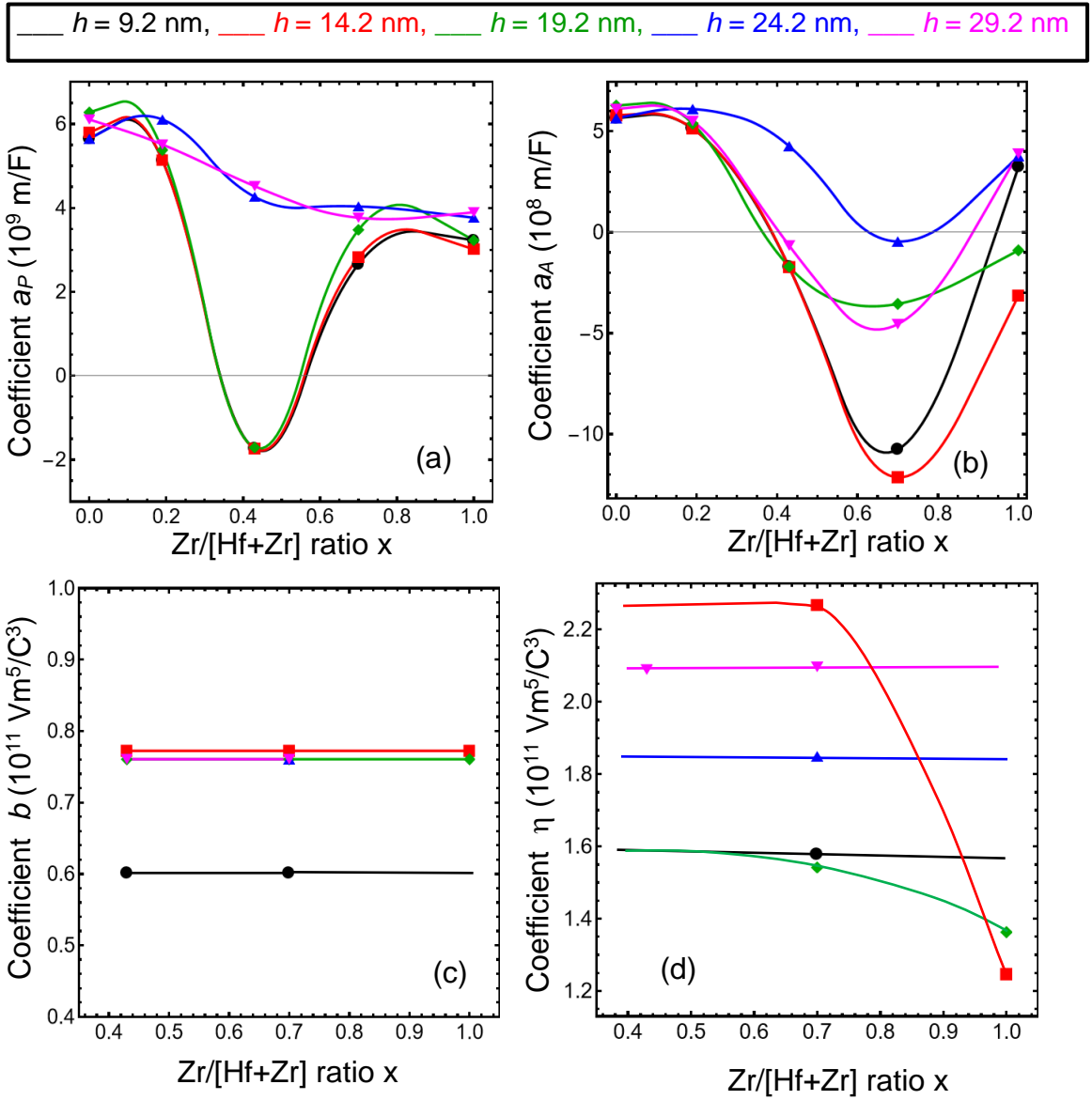

**Figure S2**. The coefficients $a_P(x,h)$ **(a)**, $a_A(x,h)$ **(b)**, $b_{P,A}(x,h)$ **(c)**, and $\eta(x,h)$ **(d)** determined from Eq.(A.7) and **Table SI**. Symbols correspond to the points extracted from Ref. [1], solid curves are the second **(a, b, d)** or first **(c)** order interpolation functions.

Noteworthy, in all our previous studies of various nanoparticles and thin films, including HfO$_2$ films, we derived or used available analytical dependences of the size- or/and thickness- dependences of the LGD expansion coefficients, because the size effect in many nanoscale ferroelectrics can be successfully described by a combination of the long-range depolarization field and strain and/or Vegard stress effects, and short-range surface and/or interface effects [3, 4, 5, 6, 7]. Disregarding our efforts, the Hf$_{1-x}$Zr$_x$O$_2$ thin films appeared to be a strongly anomalous case, as argued below in the items A)-C), which cannot be described by a conventional scheme.

A) In the most of conventional ferroelectric thin films the long-range polar or/and antipolar orders appear for $h > h_{cr}^{A,P}$, and existing analytical LGD-type models of the critical thickness $h_{cr}^{A,P}$ describe the cases. For HZO the ferroelectricity and antiferroelectricity appear in a limited range of film thicknesses, $h_{min}^P < h < h_{max}^P$ and $h_{min}^A < h < h_{max}^A$, respectively, which require major modifications of the models.

B) After a very careful analysis of the h-dependence of the coefficients α(x, h) and β(x, h) extracted directly from the experimental results [1] and shown in **Fig. S1**, it turned out that we require more experimental information about the physical nature and the concrete form of h-dependence of the LGD expansion coefficients, $a_P(x, R, h)$ and $a_A(x, R, h)$, in Eqs.(7) and how their form is dependent on the grain size, structure and the film preparation conditions.

C) Specifically, in accordance with Eqs.(7) and (8):

$$a_P(x, R, h) \cong \xi_P(x) + \frac{h_g}{\varepsilon_0(\varepsilon_b h_g + h\varepsilon_g)} + Q_{3i}(x)\left(\frac{2\gamma_i^{if}(x)}{h} + \frac{2\gamma_i^{gb}(x)}{R}\right) + Q_{3i}(x)\sigma_i(R, h), \quad \text{(A.8a)}$$

$$a_A(x, R, h) \cong \xi_A(x) + Z_{3i}(x)\left(\frac{2\gamma_i^{if}(x)}{h} + \frac{2\gamma_i^{gb}(x)}{R}\right) + Z_{3i}(x)\sigma_i(R, h). \quad \text{(A.8b)}$$

Here, the depolarization field contribution, $\frac{h_g}{\varepsilon_0(\varepsilon_b h_g + h\varepsilon_g)}$, vanishes for the case of conducting (e.g., TiN) electrodes when $h_g = 0$ (shown in **Fig. 1(b)**). The short-range surface and/or interface effects determines the terms like $\frac{2\gamma_i^{if}(x)}{h} + \frac{2\gamma_i^{gb}(x)}{R}$, or their representation via e.g., so-called extrapolation length λ. Anyway, this leads to the terms proportional to the powers of $\frac{1}{h+\lambda}$ or/and $\frac{1}{R+\lambda}$ [8]. Mismatch strains and/or quasi-uniform Vegard stresses can contribute to Eqs.(A.8) either by h-independent terms [9, 10], or by relaxation terms in the case of misfit dislocations appearance and/or surface bond contraction [8]. Note that the dependence of the average grain radius $R(h)$ is unknown and significantly depends on the film preparation conditions. Therefore, the inhomogeneous stresses, $\sigma_i(R, h)$, in Eqs.(A.8) are also unknown.

The dependences α(x, h) and β(x, h), shown in **Fig. S1**, determine the LGD expansion coefficients $a_P$ and $b_P$, in accordance with Eqs.(A.5a) and (A.5b). Thus, the analysis of the second order interpolation functions (nonmonotonic solid curves) in **Fig. S1** and **S2**, unfortunately does not allow to reconstruct the h-dependences of $R(h)$ and $\sigma_i(R, h)$, and so we were subjected to limit ourselves by the "effective" LGD model. To evolve the analytical model for the renormalized LGD coefficients of HZO one requires additional analysis of available XRD, SEM, XPS, etc., experimental results, which can provide a link between the grain structure, film thickness and preparation conditions (e.g., preparation method, annealing time, etc.).

We would like to resume the **Appendix** by the statement that the polar properties of HZO films are conditioned by the "double" size effect (when the properties are R- and h-dependent) rather than by a conventional "single" size effect (when the properties are only h-dependent). The analytical description of the "double" size effect has not been evolved yet, and we hope that the present work can be helpful in the direction.